\documentclass[tighten, times, twocolumn]{aastex62}

\received{November 2, 2018}
\revised{December 14, 2018}
\accepted{January 2, 2019}
\submitjournal{ApJ}

\shorttitle{9:7 resonance capture}
\shortauthors{Z. Cui et al.}

\begin{document}

\title{On the 9:7 Mean Motion Resonance Capture in a System of Two Equal-mass Super-Earths}

\correspondingauthor{Ewa Szuszkiewicz}
\email{szusz@feynman.fiz.univ.szczecin.pl}

\author{Zijia Cui}
\affiliation{Institute of Physics and CASA$^{\ast}$, Faculty of Mathematics and Physics, University of Szczecin, Wielkopolska 15, PL-70-451 Szczecin, Poland}

\author{John C. B. Papaloizou}
\affiliation{DAMTP, University of Cambridge, Wilberforce Road, Cambridge, CB3 0WA, UK}

\author[0000-0002-7881-2805]{Ewa Szuszkiewicz}
\affiliation{Institute of Physics and CASA$^{\ast}$, Faculty of Mathematics and Physics, University of Szczecin, Wielkopolska 15, PL-70-451 Szczecin, Poland}



\begin{abstract}

We study the formation of the 9:7 mean motion resonance in a system of 
two low-mass planets ($m_{1}=m_{2}=3M_{\oplus}$) embedded in a gaseous 
protoplanetary disk employing a full 2D hydrodynamic treatment 
of the disk-planet interactions. Our aim is to determine the disk properties
that favor a capture of two equal-mass super-Earths into this second
-order resonance. For this purpose, we have performed a series of numerical
hydrodynamic simulations of the system of two super-Earths migrating in
 disks with a variety of different initial parameters and found
 conditions for the permanent or temporary locking in the 9:7 resonance.
We observe that  capture occurs during the convergent migration of planets
if their resonance angle at the moment of arrival  at the resonance assumes
values in  a certain range (inside a window of capture).
The width of such a window depends on the relative migration and 
circularization rates that are determined by the disk parameters.
The window is wide if the relative migration rate is slow, and it becomes 
narrower as the relative migration rate increases. The window will be closed
if the migration rate is sufficiently high, and the capture will not
take place. We illustrate also how the 9:7 resonance window of capture is 
affected by the initial eccentricities and the initial orbits of the planets.

\end{abstract}



\section{Introduction} \label{sec:intro}

Mean motion resonances (MMRs) are of great importance for studying the architecture and dynamics of multiplanetary systems. Different aspects of their relevance have been discussed by many authors, for example, \cite{SzuszEdyta2012}, \cite{Fabrycky2014} and \cite{Batygin2015}. 
Commensurabilities between orbital periods have proven useful in validating planet candidates \citep{Steffen2013}.
Numerous papers have been dedicated to the first-order commensurabilities observed in such systems \citep[e.g.][]{Bryden2000, Snellgrove2001, Nelson2002, Kley2004, PapSzusz2005, Cresswell2006}. Higher-order resonances are less common than those of first order, but a number of second-order mean motion commensurabilities found recently are sufficient to arouse interest in their formation. Particularly intriguing is the system of the solar-like star Kepler-29, in which the orbital period ratio of two planets is extremely close to the nominal value for the exact 9:7 MMR \citep{Fabrycky2012}. 
More recent works have succeeded in constraining the dynamical architecture of this planetary system, performing transit timing variation analysis \citep[]{Jontof2016} and demonstrating that the planets are indeed in the resonance configuration \citep{Miga2017a}. 
The masses of the planets have been evaluated by those authors, who found them to be in the super-Earth range (a few Earth masses).

\begin{figure}[htb!]
\plotone{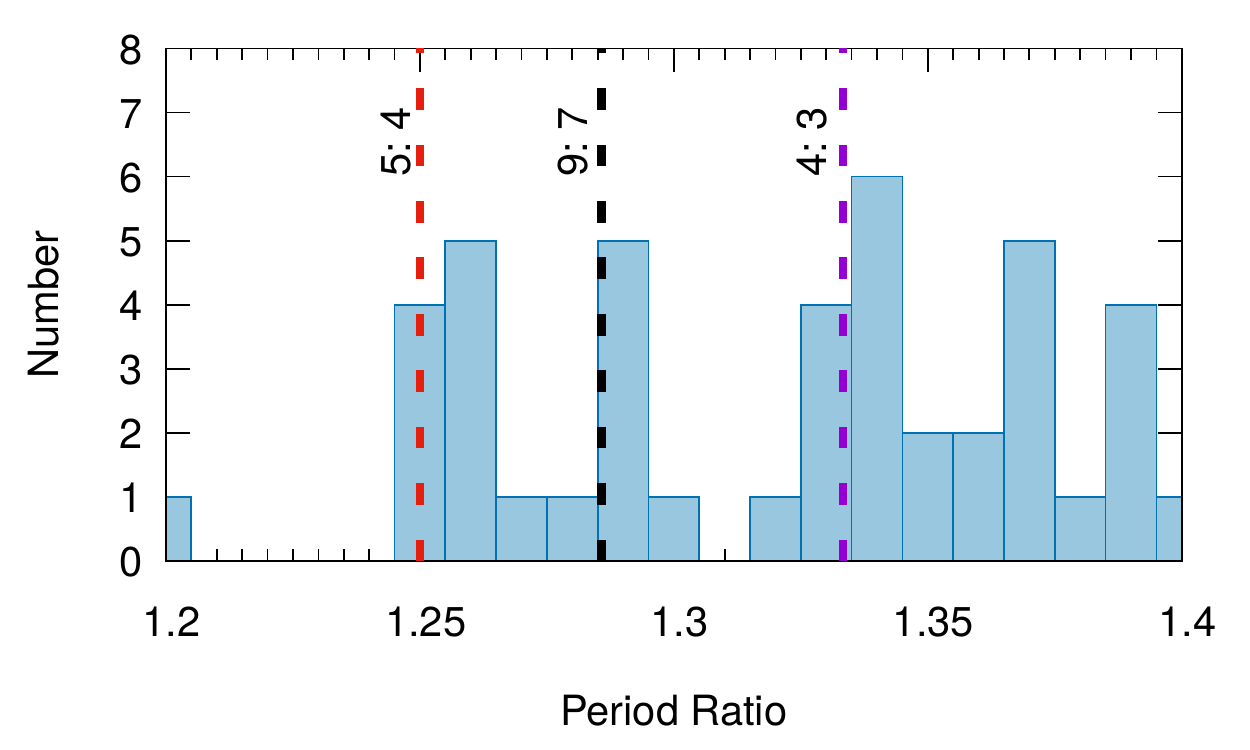}
    \caption{Histogram of confirmed {\it Kepler} planet pairs with period ratios in 
	the range  [$1.2$, $1.4$]. The width of the bin is 0.01. Dashed lines 
	show the locations of  5:4, 9:7, and 4:3 commensurabilities. 
	Data are obtained from the NASA Exoplanet Archive (as of 2018, October).}
    \label{fig:hist}
\end{figure}

As for now, the Kepler-29 system is the best example of a confirmed 9:7 resonance, but there might be other planets locked in this commensurability as well.
This is illustrated in Figure~\ref{fig:hist}, which shows the histogram of the orbital period ratios for confirmed Kepler planet pairs in the range [$1.2$, $1.4$]. 
This range of period ratios includes the 5:4,  9:7, and 4:3 resonances. The bin size is equal to 0.01 and it corresponds approximately to the mean value of the libration widths evaluated for each of these three resonances for super-Earth-mass planets. 
As can be seen from Figure~\ref{fig:hist}, there are five planet pairs with period ratios close to that of the 9:7 commensurability. We give detailed information about those planet pairs in Table~\ref{tab:nearto}.

\begin{deluxetable*}{lclccl}
\tablecaption{Planet Pairs with the Period Ratios near the 9:7 Resonance \label{tab:nearto}}
\tablecolumns{6}
\tablenum{1}
\tablewidth{0pt}
\tablehead{
\colhead{System} & \colhead{Confirmed Planet} &
\colhead{Planet Pair} & \colhead{Periods (days)} & \colhead{Deviation from 9:7} & \colhead{References} 
}
\startdata
Kepler-29 & 2 & Kepler-29b & 10.3384 & 0.0004 & \citet{Jontof2016} \\
          &   & Kepler-29c & 13.2884 &        &                    \\
\hline
Kepler-417 & 2 & Kepler-417b & 12.3309 & 0.0072 & \citet{Morton2016} \\
           &   & Kepler-417c & 15.9425 &        &                    \\
\hline
Kepler-37 & 4 & Kepler-37d & 39.7920 & 0.0009 & \citet{Hadden2014} \\
          &   & Kepler-37e & 51.1960 &        &                    \\
\hline
Kepler-1542 & 4 & Kepler-1542b & 3.9512 & 0.0053 & \citet{Morton2016} \\
          &   & Kepler-1542e & 5.1012 &        &                    \\
\hline
Kepler-33 & 5 & Kepler-33e & 31.7848 & 0.0051 & \citet{Morton2016} \\
          &   & Kepler-33f & 41.0281 &        &                    \\
\enddata
\end{deluxetable*}

Note that Kepler-29 and Kepler-417 contain two planets each, while the remaining three systems listed in the table have four (Kepler-37, Kepler-1542) or five (Kepler-33) planets orbiting around their host stars. In the Kepler-37 system 9:7 commensurability occurs between the orbital periods of planets d and e. Little is known about the planet e apart from its orbital period determined by \cite{Hadden2014}. Its status is not well established yet, as the serious doubts about its existence have been reported in \cite{Barclay2013} and \cite{Marcy2014}.
However, if the planetary nature of this object is demonstrated, the 9:7 resonance in the Kepler-37 system will be just one part of a more complex
resonance configuration with a  clear 3:1 commensurability between planets b and d. The 9:7 resonance in the Kepler-1542 system (between planets b and e) forms a chain together with the 7:6 commensurability between planets d and e. Similarly, in the Kepler-33 system the 9:7 resonance is embedded in a sequence of other commensurabilities.

These findings inspired us to perform a detailed study of 9:7 resonance capture using, for the first time, hydrodynamic simulations.
A rich body of literature on second-order resonance trapping, in the framework of both semi-analytical models and N-body calculations, has provided helpful guidelines for our investigations. \citet{Quillen2006} has explored how the probability of the capture into the 
second-order resonances depends on the strength of the resonance, migration rate, and initial particle eccentricity using a simple planar-restricted three-body Hamiltonian model. \citet{Mustill2011} have confirmed and extended that work. They found that resonance capture fails in a system with a high migration rate and has decreasing probability for higher eccentricities of the test particle. They also found that test particles can be captured by more massive planets at higher initial eccentricities and relative migration rates. \citet{Folonier2014} used an algebraic mapping of the averaged planar-restricted three-body problem to investigate the capture probability into 3:1 resonance. They found that the capture occurs for discrete windows of initial eccentricity whose locations depend on initial resonant angles, while the width of such windows is affected by relative migration rates. This indicates that the capture phenomenon is not probabilistic.

\cite{Xiang2015} study the formation of second-order resonances for planets migrating in the protoplanetary disk using numerical simulations. They used the {\it N}-body model including additional acceleration terms to calculate the evolution of two low-mass planets with wide ranges of migration parameters. From the result of surveys, they constrained the migration parameters that can induce the formation of second-order resonances. 
\citet{Xu2017} studied the capture and stability of second-order resonances for migrating planet pairs with comparable-masses and very low initial 
eccentricities by using a model based on a simplified Hamiltonian. They found that resonant capture requires slow convergent migration of the planets, with sufficiently large circularization time and small pre-resonance eccentricities. 
However, for a system with comparable masses planets and higher initial eccentricities, they conjecture that trapping should become probabilistic.
Finally, \citet{Miga2017} investigated how the formation of 9:7 resonance depends on the migration parameters and initial orbits as a result of the convergent migration of two low-mass planets by using an {\it N}-body model with migration parameters. In this work, they also gave conditions for the systems to stay in the 9:7 resonance either permanently or temporarily. This work made a successful attempt to discuss the results in relation to all above-mentioned studies, presenting in a clear way our current state of understanding on how the complex process of second-order MMR capture takes place in the early stages of the evolution of planetary systems.

In this work we take one step further and apply full 2D hydrodynamic simulations to treat the disk-planet interaction.

However, compared with other papers on disk-planet interaction that consider first-order resonances, the simulations in this work are limited to a very small region of phase space in the neighborhood of the 9:7 resonance. There are two constraining reasons for this limitation.

The first is that a very low migration rate has to be used since the time to move through the resonance must exceed the inverse libration frequency.
The latter frequency can be estimated as  \citep[see][]{Xiang2015}  
\begin{equation}
n_1-(9/7)n_2 \sim 4.2\times 10^{-5} (m_1/M_{\oplus})(M_{\odot}/M_{\star})n_2
\label{eq:intro1}
\end{equation}
\noindent leading to the constraint on the relative migration rate 
\begin{equation}
-d(a_{2}/a_{1})/dt < [4.2\times10^{-5} (m_1/M_{\oplus})(M_{\odot}/M_{\star})]^{2}n_2
\label{eq:intro2}
\end{equation}
\noindent or a migration time that exceeds $10^{6-7}$ yr.
Here $n_{1}$, $a_{1}$ and $n_{2}$, $a_{2}$ are the mean motion and semi-major axis of the inner and outer planet, respectively. $m_{1}$ and $M_{\star}$ are the masses of the inner planet and the central star, while $M_{\oplus}$ and $M_{\odot}$ are the masses of the Earth and the Sun, respectively. Such a low migration rate can only generate a very small amount of evolution in $10^{4}$ orbits, which is the typical duration of our single simulation for capture cases.

The second reason relates to eccentricity damping. As a second-order resonance trapping in a nonadiabatic regime, as considered in this work, 
requires nonzero eccentricity \citep{Folonier2014}, the initial eccentricity must survive the journey to resonance from outside against circularization. With no source of orbital eccentricity, the system can only migrate for a time comparable to the circularization time.

For the reasons mentioned above, the work in this paper is restricted to looking only at the last stages of capture into the resonance that
establishes libration and locked evolution. Thus, it is very much a local treatment.

The plan of this paper is as follows. In Section~\ref{sec:method} we describe our methods. Section~\ref{sec:result} contains the results of a survey of the outcomes of the migration of pairs of two equal-mass super-Earths for the relevant set of the initial conditions that can lead to capture in the 9:7 resonance.
The conclusions and discussion of our findings are presented in Section~\ref{sec:conclusion}.


\section{Disk and Planet Parameters Adopted for Our Investigations} \label{sec:method}

We consider a system of two planets with masses $m_{1}$ and $m_{2}$ embedded in a gaseous protoplanetary disk and orbiting a central star with the mass of $M_{\star}$. A typical protoplanetary disk is to a good approximation geometrically thin and rotates practically with the Keplerian angular velocity $\Omega$. For this reason, we have chosen to work in the framework of the 2D vertically integrated disk model, using cylindrical coordinates ($r, \phi, z$). The origin of the coordinate system is located at the position of the central star. 
The gas in the disk is modeled by adopting a locally isothermal equation of state, which means that the vertically integrated pressure $P$ and density $\Sigma$ satisfy the relation $P=\Sigma c_{s}^{2}$, where $c_{s}$ is the sound speed. The sound speed is related to the vertical scale height of the disk $H$ as follows:  $H=c_{s}/\Omega$. Assuming that the aspect ratio $h=H/r$ is constant in the region of planet formation, as has been argued by \cite{Ruden86} and \cite{Terquem00}, we obtain the fixed temperature profile proportional to $r^{-1}$. 
Our choice of a simple locally isothermal equation of state can be justified on the basis of results presented by \citet{Kley2008}. 
These authors  made a comparison between the torques acting on a planet in fully radiative and locally isothermal disks. They found that for the low-mass planets considered in our calculations, the torques acting in these cases were similar. The implication is that the generic form of the migration rates is not significantly affected by our particular choice of  equation of state. The disk self-gravity is neglected since the masses of the disks considered in this work are sufficiently low.

The system of units adopted in the simulations is that the unit of mass is the mass of the central star $M_{\star}.$ The unit of length is the initial orbital radius of the inner planet, $r_{1},$ and the unit of time is the initial orbital period of the inner planet $P_{1}.$ In this work, we initiate the inner planet at $r_{1}=1$ {\it au} and take $M_{\star}$ to be a solar mass. Then, the time unit for the numerical simulations is $1 $ yr. 

The eccentricities of the inner- and outer-planet orbits in our calculations, if not stated otherwise, are respectively set to be $e_{1}=0.005$ and $e_{2}=0$. Note that the inner planet is put at the pericenter of its orbit, which means that the initial semi-major axis of the inner planet 
is $a_{1}=r_{1}/(1-e_{1})$. The initial value of the orbital radius of the outer planet, $r_{2},$ is chosen such that the initial orbital period ratio of the planets is slightly larger than 9:7, usually 1.2865.

Torques produced by disk-planet interaction determine the migration rate of a planet. In the linear regime the migration rate is proportional to the planet mass and the disk surface density \citep[e.g.][]{Tanaka2002}. We consider a pair of planets of equal small mass. We adopt an initial surface density profile that is uniform in the outer part of the disk with a slope causing it to decrease while moving into the inner part of the disk. The initial surface density profile adopted, $\Sigma(r)$, is given by
\begin{eqnarray}
\label{disk}
\Sigma(r) = \Sigma_0 r \hspace{5.5mm} {\rm for} & r_{\rm min} < r < 1.1, \nonumber  \\
\Sigma(r) = 1.1\Sigma_0 \hspace{3mm} {\rm for } & 1.1 \leq r < r_{\rm max},
\end{eqnarray}
where $\Sigma_{0}$ is a scaling parameter, with $r_{\rm min}$ and $r_{\rm max}$ being the inner and outer bounding radii of our computational domain, respectively. This particular surface density profile has been found to guarantee the convergent migration of low equal-mass planets in the early stages of the evolution. We insert the inner planet in the region of the surface density profile with a positive gradient and the outer planet in the flat part of the profile. In this way, on account of corotation torques \citep[][]{Paarde2009}, the inner planet's migration is slowed down relative to that of the outer planet, ensuring that the planets will have convergent migration. This is a condition for the formation of 
MMRs \citep[e.g.,][]{Nelson2002, Kley2004}. 

We remark that a region of the disk where the surface density starts to decrease inward is expected at the boundary of the region interior to which there is turbulence produced by the magnetorotational instability throughout the vertical extent of the disk \citep[see e.g.] [and references therein]{Andre2016}. There is expected to be an increase in effective viscosity as this inner region is entered. However, noting that the planets in our simulations migrate over a small radial extent on the order of the vertical thickness and are in the type I migration regime, we have not attempted to model this in any detail here.

In this work, we use the hydrodynamic code NIRVANA\citep[e.g.,][]{Ziegler1998} to solve the Navier-Stokes equations in order to calculate the evolution of the disk. The action of an effective viscosity resulting from turbulence is modeled by adopting a constant kinematic viscosity $\nu.$
Details of the numerical scheme can be found in \citet{Nelson2000}. To define the computational domain, we select $r=r_{min}=0.33$ and $r=r_{max}=3.0$ as open radial boundaries and azimuthal domain given by $\phi \in [0, 2\pi].$ The choice of  numerical resolution has been preceded by a series of convergence tests that show that for a doubling of the resolution finally adopted in each coordinate direction the circularization rates remain the same. On the other hand, the relative migration rate, sometimes being determined by a near cancellation between the rates for the individual planets, may change by up to a factor of two. However, this was found to have the effect of mapping simulation results to those that would apply to different disk parameters at lower resolution; this fact will not change the final conclusions of this work, which relate to  structures revealed by an ensemble of simulations. In consequence, as a compromise between specific numerical accuracy and improved statistical significance of our results, the computational domain was uniformly divided into 384 cells in radius and 512 cells in azimuth.

The gravitational potential of the planets is smoothed by the incorporation of a softening parameter, $b$, equal to $0.45H$. We consider the system in which both planets' masses are in the super-Earth range. For the disk aspect ratio,  $h=H/r=0.045,$ adopted in this work, the orbital migration of the two planets is in the type I regime. 


\section{Two Equal-mass Super-Earths Migrating in a Protoplanetary Disk} \label{sec:result}

In this section we describe the evolution of two equal-mass planets ($m_{1}=m_{2}=3M_{\oplus}$) evolving dynamically in a gaseous protoplanetary disk in the vicinity of the 9:7 MMR using full 2D hydrodynamic simulations. The main aim of this investigation is to determine physical conditions that can lead to the formation of the 9:7 commensurability.
In our approach we take advantage of the explicit treatment of the disk-planet interactions in the hydrodynamic code. Whenever possible we compare our results with those obtained by analytic methods and {\it N}-body calculations performed recently by other authors.
In order to achieve our aim, we follow the evolution of planets in  disks with a given specification of the initial surface density scaling parameter $\Sigma_{0}$ and viscosity $\nu$ until the planets reach the 9:7 resonance. The outcome of this evolution is a capture in or a passage through this commensurability. We do not continue our simulation for long enough to be able to determine whether the capture is permanent or not.
In consequence, the results of our survey indicate the range of disk parameters for which the planets will be locked (at least temporary) in the 9:7 resonance.
 

\subsection{A 9:7 Resonance Survey} \label{sec:survey}

We start our calculations from a planet configuration, such that the initial orbital period ratio of the planets is slightly larger than 9:7, namely, about 1.2865. We do not consider here how the planets arrived at these positions, but simply assume that such a configuration has formed during the previous stage of evolution.

\begin{figure*}[htb!]
\includegraphics[height=7cm,width=1.0\columnwidth]{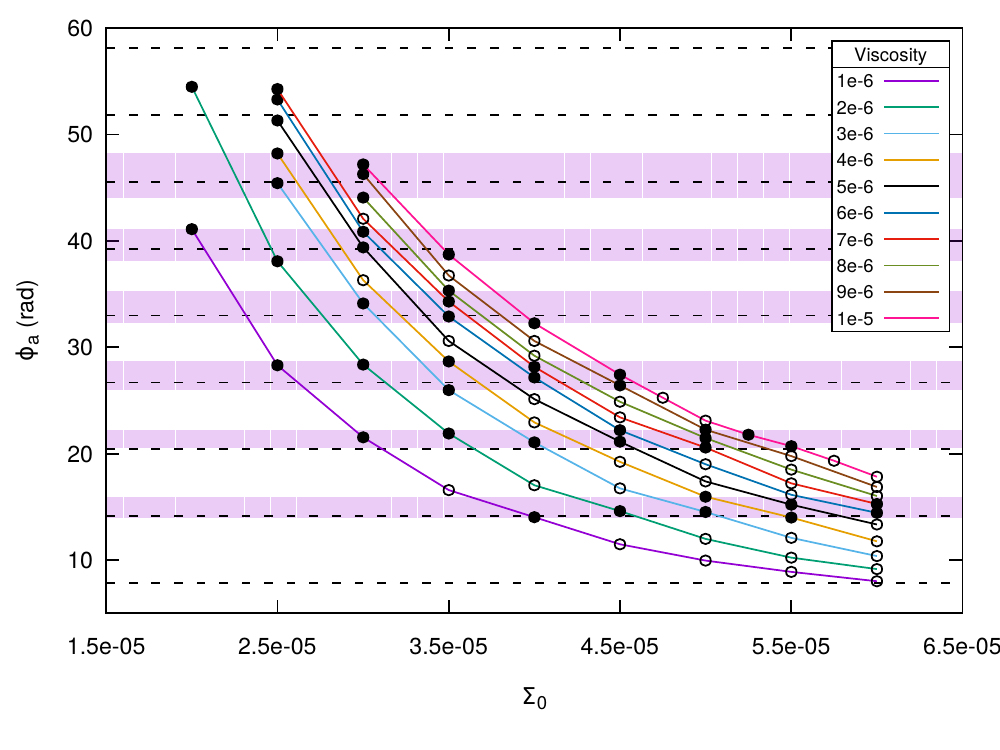}
\includegraphics[height=7cm,width=1.0\columnwidth]{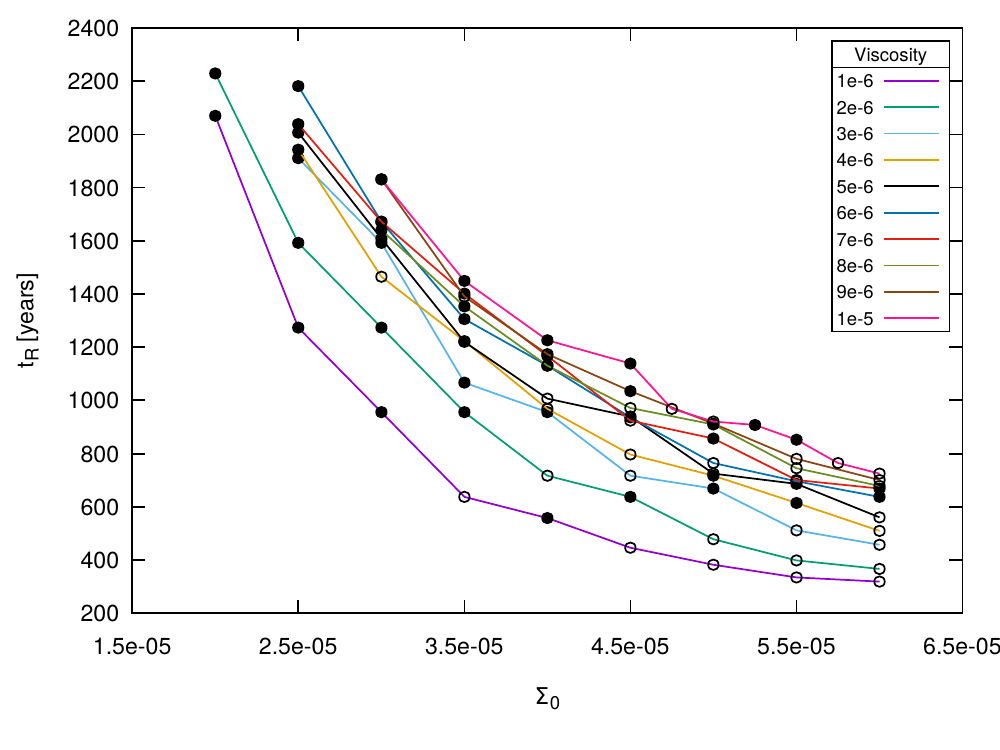}
\caption{Left panel: accumulated resonance angle $\phi_{a}$ as a function of the surface density parameter $\Sigma_0$ for various values of the disk viscosity $\nu$. The filled black circles indicate those simulations in which the planets become locked in the 9:7 resonance, and the open circles represent those cases for which the planets passed through the commensurability. The dashed lines indicate the positions of $\phi_{a}$ with $\phi_{\rm entry}=\frac{3}{2}\pi$. The violet regions (stripes) show the ranges of $\phi_{a}$ for which the capture into the resonance can occur. Right panel: results of the same simulations as in the left panel, but with the ordinate considered to be the time $t_{R}$ taken between the start of the simulation and the point at which the resonance was entered expressed as orbital periods at the initial location of the inner planet (1 yr for 1 au).
\label{fig:ecc10_map}}
\end{figure*}

We calculate the evolution of planets in protoplanetary disks with the initial surface density scaling parameter, $\Sigma_0,$ chosen to be in the interval $[2 \times 10^{-5}, 6 \times 10^{-5}]$ and kinematic viscosity in the interval $[10^{-6}, 10^{-5}]$ in units of $M_{\star}/r_{1}^{2}$ and $r_{1}^{2}(GM_{\star}/r_{1}^{3})^{1/2}$, respectively. The results from $82$ simulations are illustrated in Figure~\ref{fig:ecc10_map} in the form of
a map of the simulation outcomes, where the black filled circles denote ``a capture by'' and open circles ``a passage through'' the 9:7 resonance, depending on the disk parameters $\Sigma_0$ (horizontal axis) and $\nu$.
Lines of an indicated color connect filled and open circles corresponding to a fixed value of $\nu$. There is a clear regularity in the occurrence of the resonance captures seen in this figure, and it can be conveniently expressed in terms of a new quantity, which we call an accumulated resonance angle $\phi_{a}$, and which is specified on the vertical axis of our map. This quantity is constructed using the resonance angle $\phi_{1}$ defined as follows:
\begin{equation}
  \phi_{1} = 9\lambda_{2} - 7\lambda_{1} -2\varpi_{1}
  \label{eq:phi1}
\end{equation}
where $\lambda_{1}$ and $\lambda_{2}$ are the mean longitudes of the inner and outer planets, respectively, while $\varpi_{1}$ is the longitude of the pericenter of the inner planet. We comment that the two resonance angles $\phi_{2}$ and $\phi_{3}$ being defined through
\begin{equation}
  \phi_{2} = 9\lambda_{2} - 7\lambda_{1} -2\varpi_{2} \quad{\rm and}\quad
  \phi_{3} = 9\lambda_{2} - 7\lambda_{1} -\varpi_{1} - \varpi_{2}. 
  \label{eq:phi1}
\end{equation}
are related to $\phi_1$ by subtracting multiples of the angle between the apsidal lines of the planets, $\Delta \varpi= \varpi_1 - \varpi_2,$ with $\varpi_2$ being the longitude of pericenter of the outer planet.

The meaning of $\phi_{a}$ is easy to understand looking at the variation of the $\phi_{1}$ illustrated in Figure~\ref{fig:acc_angle} for the case of two super-Earths evolving in a disk with $\Sigma_{0}=4.5 \times 10^{-5}$ and $\nu= 2 \times 10^{-6}$. At the beginning of the simulation the resonance angle performs a complete rotation from zero to $2\pi$ twice, accordingly, the number of full rotations $N_{\rm rot}$ performed by the resonance angle is equal to 2. As the calculations proceed, $\phi_{1}$  reaches the value of about 4.2 rad at the moment corresponding to when planets enter the 9:7 resonance. We denote the value of $\phi_{1}$ at this particular moment of time as $\phi_{\rm entry}$. Now, the accumulated resonance angle $\phi_{a}$ can be calculated as
\begin{equation}
 \phi_{a} = 2\pi N_{\rm rot} + (2\pi - \phi_{\rm entry})= 14.65 \ {\rm rad}.
 \label{eq:accum}
 \end{equation}
Before discussing the results presented in the map, we make some remarks concerning the determination of the time of arrival at the resonance.
The resonant angle, $\phi_{1}$, starts to librate when the period ratio is close but not equal to the value implied by strict commensurability
due to the finite width of the resonance where libration can occur. As already mentioned in Section \ref{sec:intro} in connection with Equation (\ref{eq:intro1}), {\citet{Xiang2015} estimate that for small eccentricities the entrance into resonance occurs when
\begin{equation}
|n_1/n_2 - 9/7| \sim 4.2\times 10^{-5}(\frac{m_1}{M_{\oplus}})(\frac{M_{\odot}}{M_{\star}}).
\label{X11}
\end{equation}
A corresponding expression for larger eccentricities \\
  $ e_1 > \sim 2\times 10^{-3} \left[(m_1/M_{\oplus})(M_{\odot}/M_{\star})\right]^{1/2} $ is
\begin{equation}
|n_1/n_2 - 9/7| \sim 2.6\times10^{-3}(\frac{e_{1}}{0.1})\left[(\frac{m_1}{M_{\oplus}})(\frac{M_{\odot}}{M_{\star}})\right]^{\frac{1}{2}}.
\label{X22}
\end{equation}
Using Equations (\ref{X11}) and (\ref{X22}), the time for arriving at the resonance can be determined and the accumulated resonant angle can be measured.

For the case of two equal-mass planets with  small eccentricities $e_1$ less than $\sim 0.003$, we find from Equation (\ref{X11}) that $\phi_{1}$ starts to librate when the period ratio is close to 1.2858. This is exactly the moment of time indicated in Figure~\ref{fig:acc_angle} by the vertical green dot-dashed line, which determines the value of $\phi_{\rm entry}$. In this way we obtain one of the points of our map illustrated in Figure~\ref{fig:ecc10_map}, namely, the one that is located in the first violet stripe, counting from the bottom of the figure for $\Sigma_0 =4.5\times 10^{-5}$. 

\begin{figure}[htb!]
\plotone{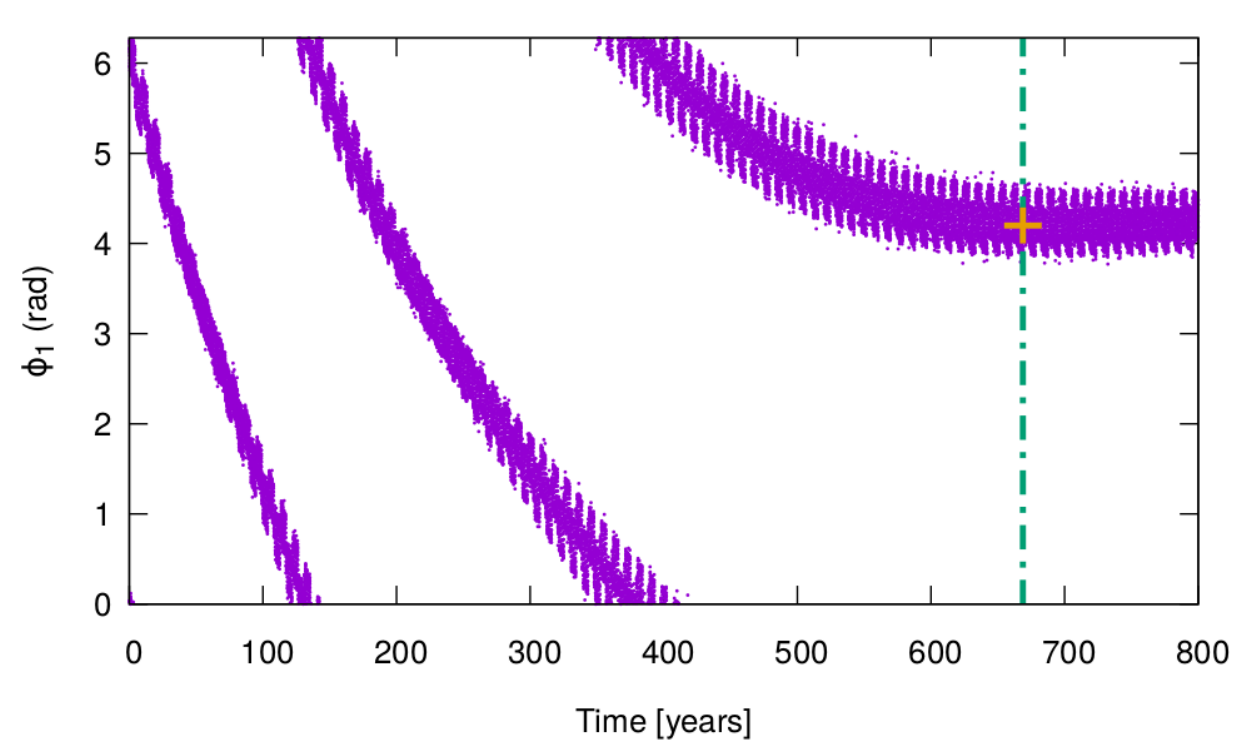}
\caption{Evolution of a resonant angle $\phi_{1}$ from the beginning of the calculation
          until the moment of the planet entry to the 9:7 resonance. The green dot-dashed line indicates the
          time when the period ratio is equal to 1.2858. The plus sign represents the value of
          $\phi_{\rm entry}$ in this simulation. The values of $\Sigma_0$ and $\nu$ were $4.5\times 10^{-5}$
          and $2\times 10^{-6}$, respectively.}
\label{fig:acc_angle}
\end{figure}

\begin{figure}[htb!]
\plotone{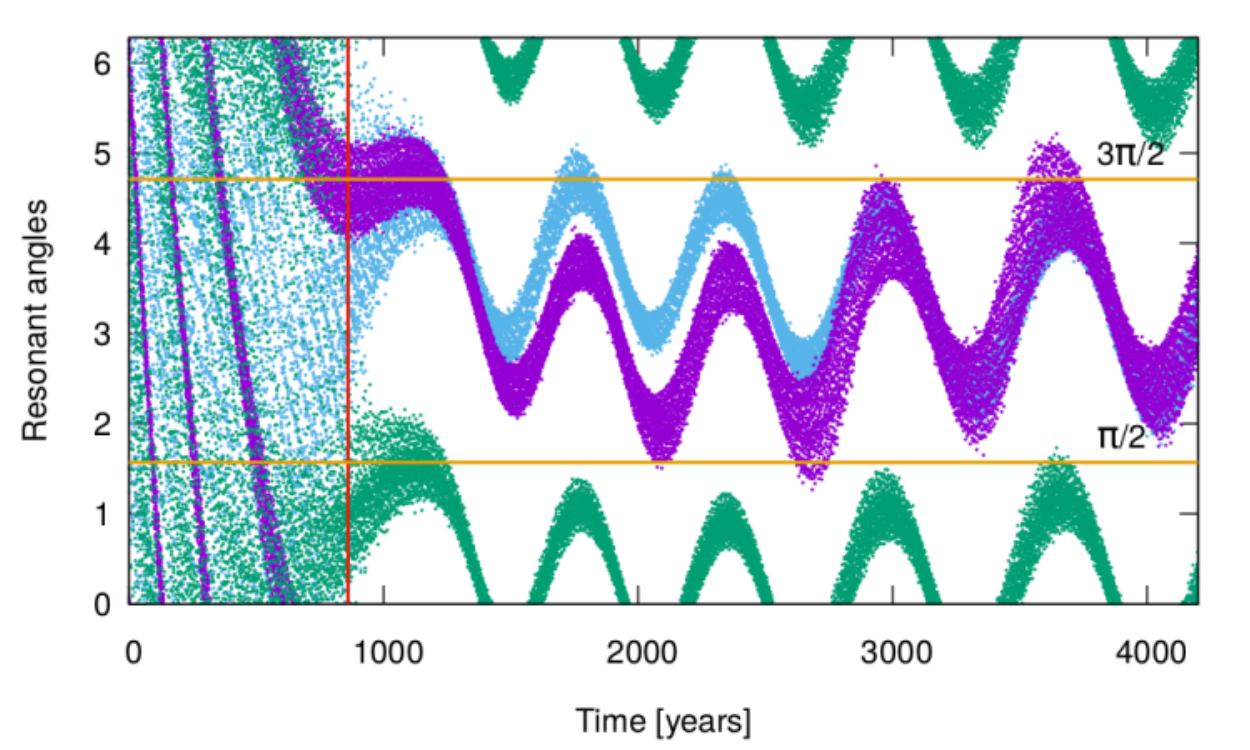}
\caption{Evolution of the resonant angles $\phi_{1}$ (violet), $\phi_2$ (blue), and $\phi_3$ (green) as a function of time measured in orbital periods at the initial location of the inner planet (1 yr for 1 au). The values of $\Sigma_0$ and $\nu$ for this simulation  were $5\times 10^{-5}$ and $7\times 10^{-6}$ respectively. The red vertical line indicates the time at which the planets arrive at the upper boundary of the 9:7 resonance width measured from the beginning of the calculation. At this time the period ratio is equal to 1.2858 and $\phi_1$ is librating around $\frac{3}{2}\pi$. However, at later times the libration becomes centered around $\pi.$}
\label{fig:acc_angleslib}
\end{figure}

To further illustrate the above points, we present results from a simulation for which resonance ``capture'' takes place and for which the values of $\Sigma_0$ and $\nu$ were $5\times 10^{-5}$ and $7\times 10^{-6}$ respectively. The time evolution of the resonance angles is shown in Figure \ref{fig:acc_angleslib}. The red vertical line indicates the time at which the planets arrive at the upper boundary of the 9:7 resonance width.
At this time the period ratio is equal to 1.2858 and $\phi_1$ is librating around $3\pi/2$, which is accordingly identified as $\phi_{\rm entry}.$
After further evolution, the libration, albeit of large amplitude, becomes centered around $\pi$. Thus, it is important to note that in general the librations of $\phi_1$ will not necessarily be centered on $\phi_{\rm entry}$. From Figure \ref{fig:acc_angleslib}, we also can see that when capture into the resonance occurs the resonant angles $\phi_2$ and $\phi_3$ are also found to enter into libration.

Changing the disk parameters (the surface density scaling  and viscosity), we calculate $\phi_{a}$ for all our simulations and plot our results on the map in Figure \ref{fig:ecc10_map}. The black filled circles are for the simulations in which the planets are locked in the resonance (``capture'' cases), and the open circles represent the cases for which libration of $\phi_1$ and locking into the resonance did not occur (``fail'' cases). 
We remark that $\phi_{a}$ was determined by evaluating it when the period ratio attained the value of 1.2858 for both the ``fail'' cases and the ``capture'' cases. It is seen that the values of $\phi_{a}$ for which there was ``capture'' are distributed in particular regions of the $(\Sigma_0,  \phi_a)$ plane, which are indicated by violet stripes. 
The 'capture' cases in the same violet stripe region have $\phi_{a}$ with the same value of $N_{\rm rot}$ (from bottom to top, $N_{\rm rot}$ is increasing from 2 to 7). In the figure we do not draw the violet stripe for the simulations with $N_{\rm rot}=8$ in the range of $\phi_{a}>50$ since the number of the simulations in this range is not enough to infer the width of this 'capture' region.

Moreover, these regions either contain those $\phi_{a}$ for which $\phi_{\rm entry} = \frac{3}{2}\pi$ {or have them lying on a boundary}. This means that `entry' into the resonance happens most likely for a value of $\phi_{\rm entry}$ close to $\frac{3}{2}\pi$. However, it is important to note, given the limited number of simulations performed, that we cannot infer that ``capture'' is certain in these regions, only that it appears to be more likely. If the number of simulations was to be increased and ``capture'' was probabilistic, these regions could not remain as domains where it could be stated that capture was inevitable.

It is also clear from Figure~\ref{fig:ecc10_map} that for higher $\phi_{a}$ the capture regions could become wider, although small number statistics could prevent us from seeing that they could become more fragmented. If they did not become so, and the capture regions continue to increase in width with capture remaining certain within them, for large enough $\phi_{a}$, the capture regions could cover the full $2\pi$, which means that
the capture probability becomes $100\%$. This corresponds to orbital evolution with a very low relative migration rate. In the opposite limit we do not find any capture regions below $\phi_{a}= 13.33$. In this case the relative migration rate of the planet is high and the probability of capture is rather low or even zero. However, we cannot exclude the existence of very narrow capture regions there owing to the limitations in the sampling procedure used in this survey. 

The map shown in Figure~\ref{fig:ecc10_map} allows one to predict the results of simulations in terms of resonance capture for given values of accumulated resonant angle, which can be translated into $\Sigma_{0}$ and $\nu$ taken from the relevant parameter space. In other words, we can foresee whether the planets will have a significant chance of becoming locked into the 9:7 resonance or not for the disk parameters of interest.

\begin{figure}[htb!]
\centering
\includegraphics[width=\columnwidth]{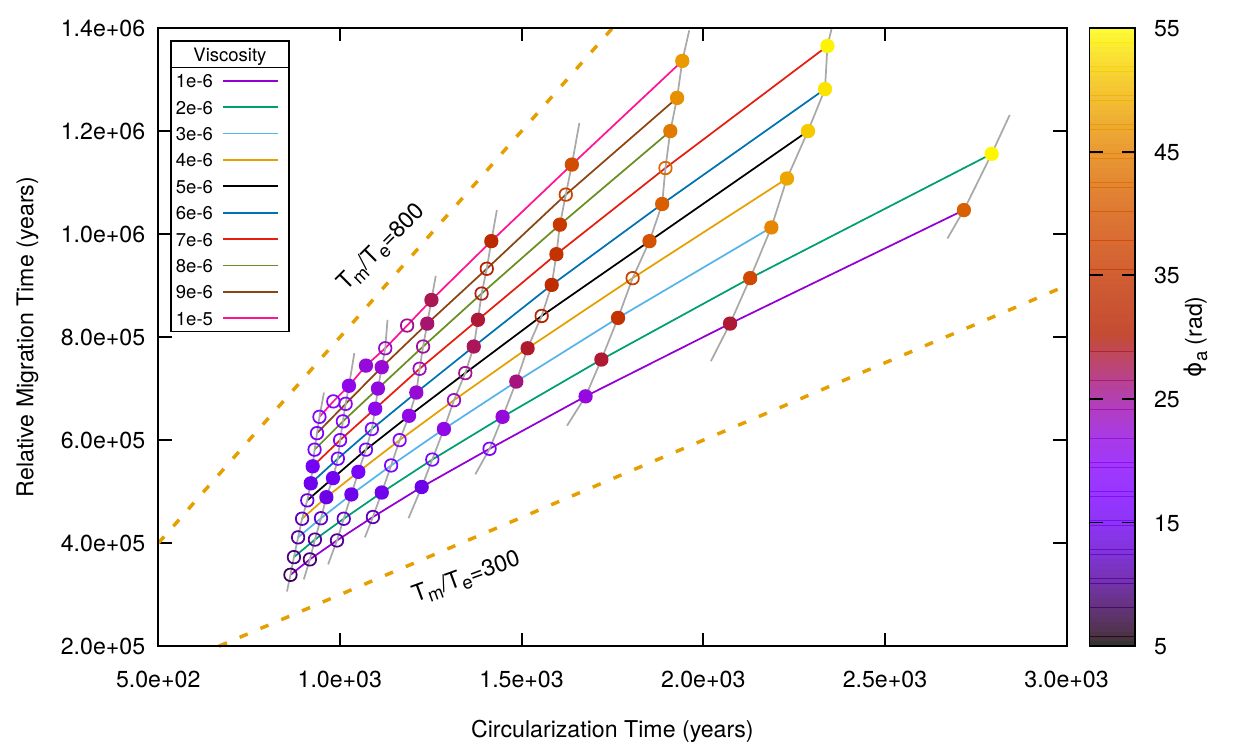}
\caption{Value of $\phi_{a}$, the circularization time $T_{e}= - e_1/{\dot e_1}$ and the relative migration time $T_{m}= a/\dot{a}$ for the simulations plotted in the map  shown in Figure~\ref{fig:ecc10_map}. The value of the kinematic viscosity adopted in the simulations is indicated by the color of the curves on which representative points lie. Results connected by gray curves are from simulations with the same surface density scaling parameter. The values of this are the same as in Figure~\ref{fig:ecc10_map}, thus lying in the interval $[2 \times 10^{-5}, 6 \times 10^{-5}]$ (increasing from right to left).}
\label{fig:ecc10_contour_refit}
\end{figure} 

The right panel of Figure~\ref{fig:ecc10_map} plots the results presented in the left panel in the $(\Sigma_0, t_R)$ plane. Thus, we replace $\phi_a$ by the time, $t_R ,$ at entry into the 9:7 resonance. As expected, the plot looks qualitatively similar; however, although there is a marked tendency for the time of resonance entry to increase with viscosity, there is not always a clear monotonic increase as is the case with $\phi_a$. Furthermore, there is no information that can be discerned about resonant angles and domains of ``capture'' appear distorted. 

The results presented in Figure~\ref{fig:ecc10_map} can also be discussed in terms of physical quantities that determine the evolution of planets in a protoplanetary disk due to disk-planet interactions, namely, the relative migration rate and the circularization rate. These are defined as $\dot{a}/a=\dot{a_{1}}/a_{1}-\dot{a_{2}}/a_{2}$ and $- \dot{e_{1}}/e_{1}$ respectively, where $a_{1}$, $a_{2}$ are the semi-major axes of the inner and outer planets and $a=a_{2}/a_{1}$.

In our analysis we adopt the values of these quantities derived from the early orbital evolution of the system in [0, 160] yr. It is important to specify the period of time, which we have considered because as the evolution proceeds the relative migration and circularization rates change gradually with time.  
The disk parameters ($\Sigma_0$, $\nu$) determine the initial relative migration rate of the planets, $\dot{a}/a$, and the circularization rate $-\dot{e_{1}}/e_1,$ so the outcome of our simulations can be expressed also in terms of these quantities. In order to make the comparison with Figure~\ref{fig:ecc10_map} easier, we use as ordinate the quantity $T_{m}=a/\dot{a}$, instead of the relative migration rate and as abscissa the circularization time $T_{e}=-e_{1}/\dot{e_{1}}$, instead of the circularization rate.
The interconnection between the disk parameters, relative migration time, circularization time, and the value of $\phi_{a}$ is illustrated in
Figure~\ref{fig:ecc10_contour_refit}.

The values of the kinematic viscosity are indicated by curves of different color as in previous figures. The points plotted on gray curves show the results of simulations with the same surface density scaling parameter. From the right to left, the value of $\Sigma_0$ is increasing while remaining in the range of $[2 \times 10^{-5}, 6 \times 10^{-5}]$.

\begin{figure*}[htb!]
\plottwo{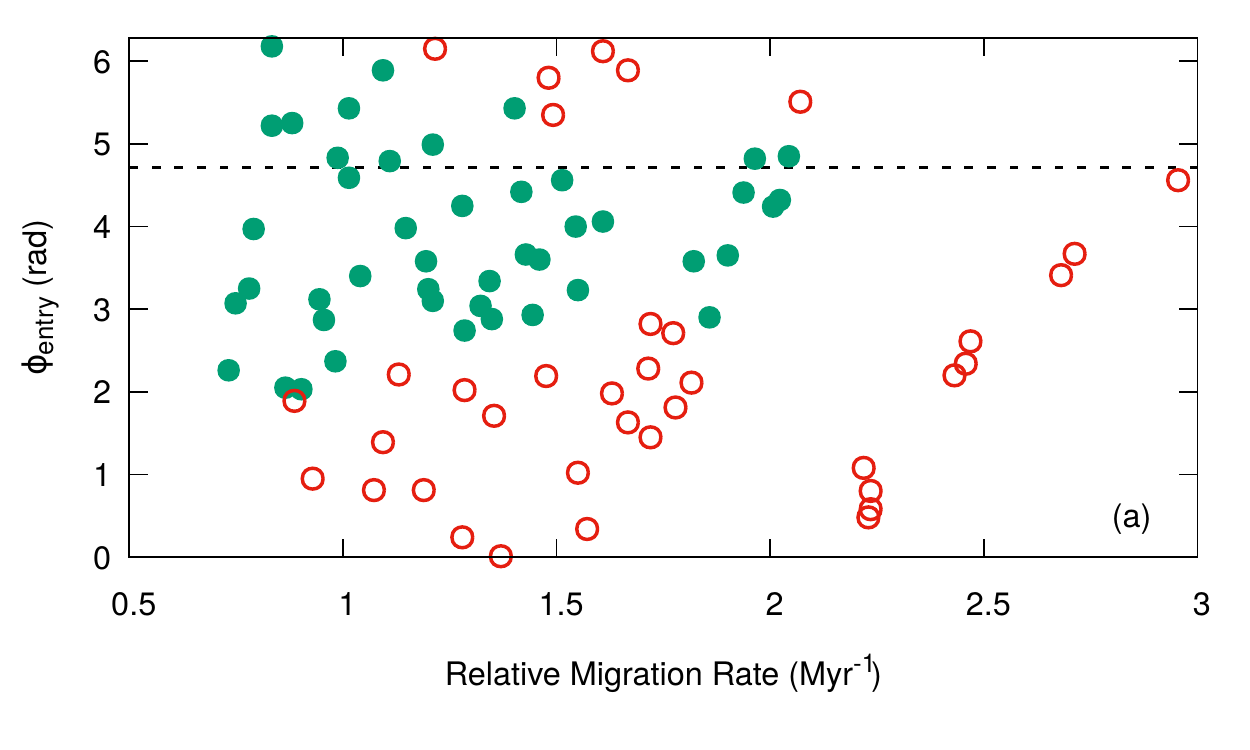}{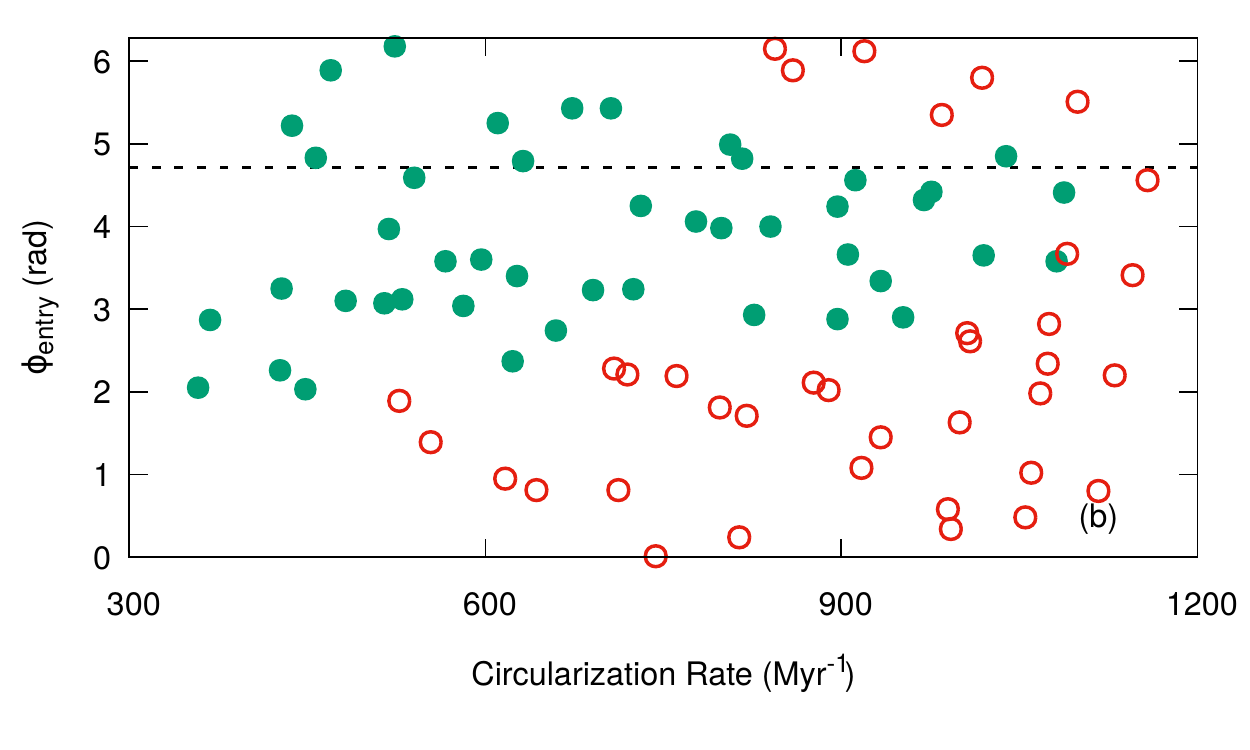}
\caption{Values of $\phi_{\rm entry}$ as a function of relative migration rate (left) and circularization rate (right) are respectively illustrated in the left and right panels for simulations used to construct the map shown in Figure~\ref{fig:ecc10_map}. The green filled circles indicate cases where ``capture'' occurred, while the red open circles represent ``fail'' cases for which ``capture'' did not occur. The position of $\frac{3}{2}\pi$ is shown by a dashed horizontal line in two figures.}
\label{fig:dot_angle}
\end{figure*}

As seen in Figure~\ref{fig:ecc10_contour_refit} with higher surface density and lower viscosity, the evolution of planets has larger $-\dot{e_{1}}/e_{1}$ and $\dot{a}/a$ (or smaller $T_{e}$ and $T_{m}$), and vice versa. For simulations with $T_{m} < 0.48$ Myr, which means that the relative migration rate is $> 2.07 ~{\rm Myr}^{-1}$, all simulations result in passage through the 9:7 resonance. 
However,  when $T_{m} \in [0.49,0.55] $ Myr, we obtain eight ``capture'' cases with $2.8 < \phi_{\rm entry} < 4.8$, which is shown as the first region of $\phi_{a}$ for capture at the bottom of Figure~\ref{fig:ecc10_map}. In this way, other wider ``capture'' regions can be found in the regime of low relative migration rates and circularization rates. We observe that the occurrence of 9:7 resonance does not depend only on $T_{m}$ or $T_{e}$ but on both of them as indicated by the distribution of ``capture'' and ``fail'' cases in Figure~\ref{fig:ecc10_contour_refit}. 

The relations of $\phi_{\rm entry}$ to the relative migration rate and the circularization rate that play an important role in the map simulations are respectively shown in Figure~\ref{fig:dot_angle} in two separate panels. The value $\phi_{\rm entry}=\frac{3}{2}\pi$ is indicated by a dashed horizontal line in each panel. We found that for specific values for the  relative migration rate or circularization rate, the values of $\phi_{\rm entry}$ for ``capture'' cases gather in a particular domain. The width of this domain is narrower for simulations with higher relative migration rate, as can be seen in the left hand panel of Figure~\ref{fig:dot_angle}. Since $\dot{a}/a$ determines the time of entry into the 9:7 resonance, the value of $\phi_{a}$ is affected by this quantity directly. This results in  $\phi_{a}$ being larger in simulations  with low relative migration rate. It is seen from both panels of Figure~\ref{fig:dot_angle} that the domain for capture decreases in width moving from left to right.

A similar trend can also be seen in the right hand panel of Figure~\ref{fig:dot_angle}, in which the domain of ``capture'' becomes smaller as the  circularization rate is increased. We infer that capture into the 9:7 resonance can occur with $\phi_{\rm entry}$ in a region whose range should depend on both $\dot{a}/a$ and $-\dot{e_{1}}/e_{1}$. 


\subsection{The Formation of the 9:7 Resonance} \label{sec:formation}

\begin{figure*}[htb!]
\centerline{
\vbox{
\hbox{
\includegraphics[width=0.25\textwidth]{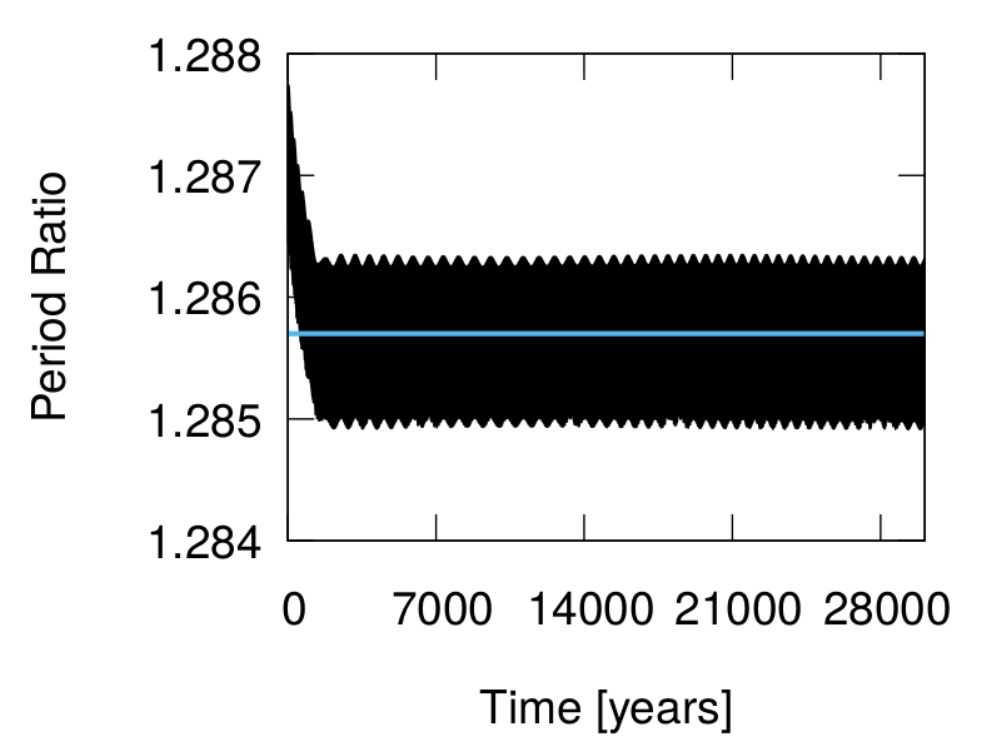}
\includegraphics[width=0.25\textwidth]{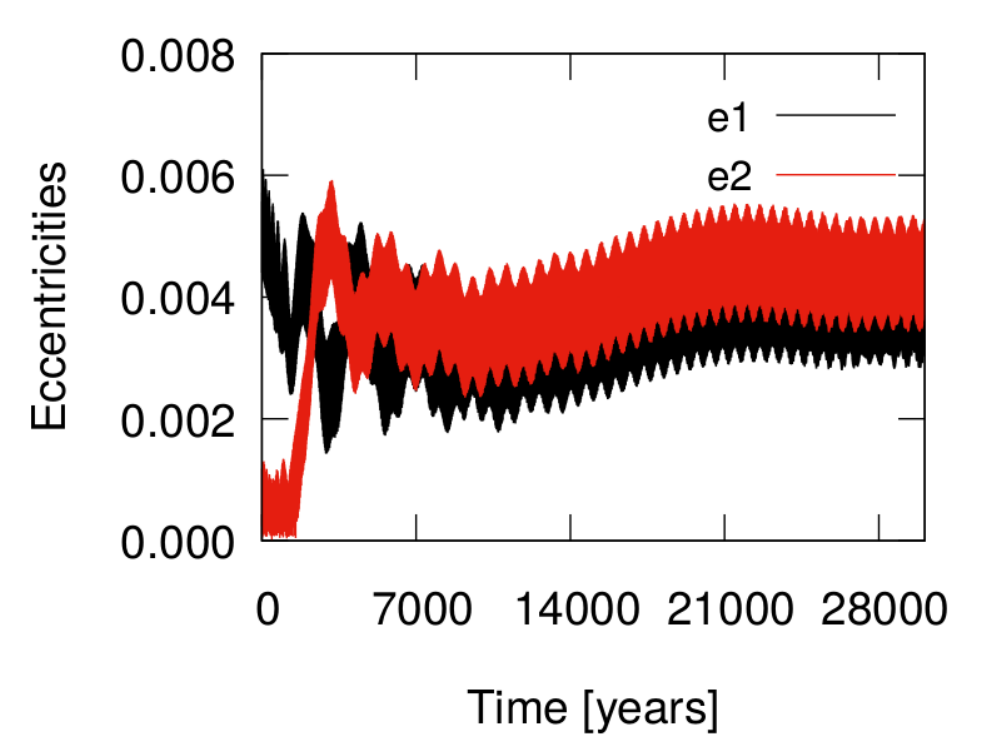}
\includegraphics[width=0.25\textwidth]{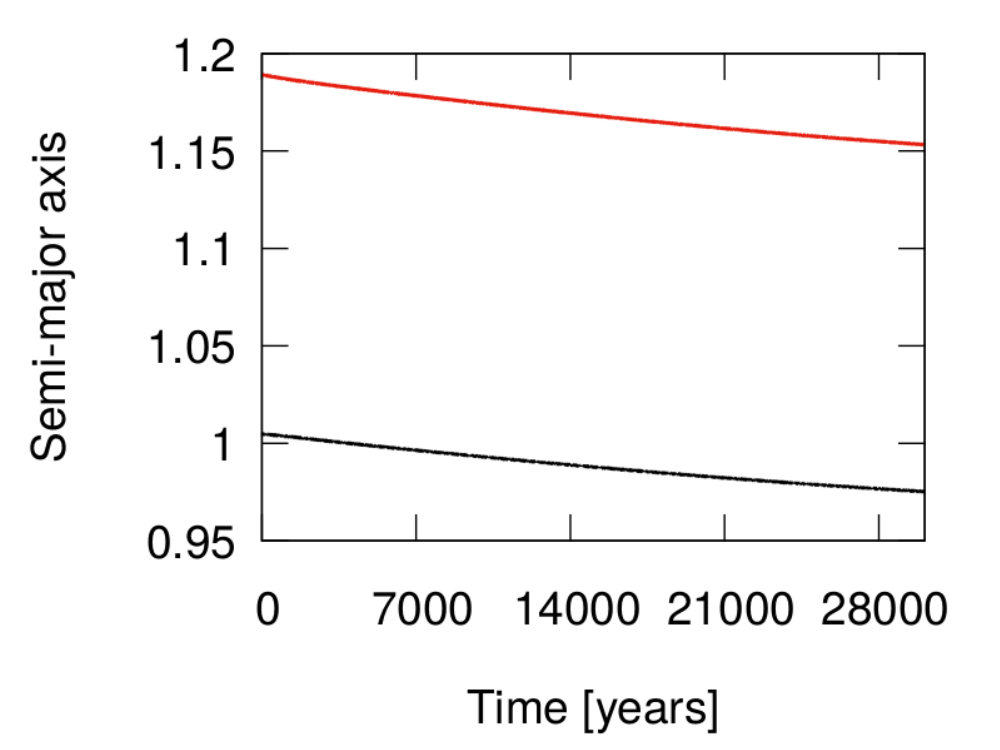}
}
\hbox{
\includegraphics[width=0.25\textwidth]{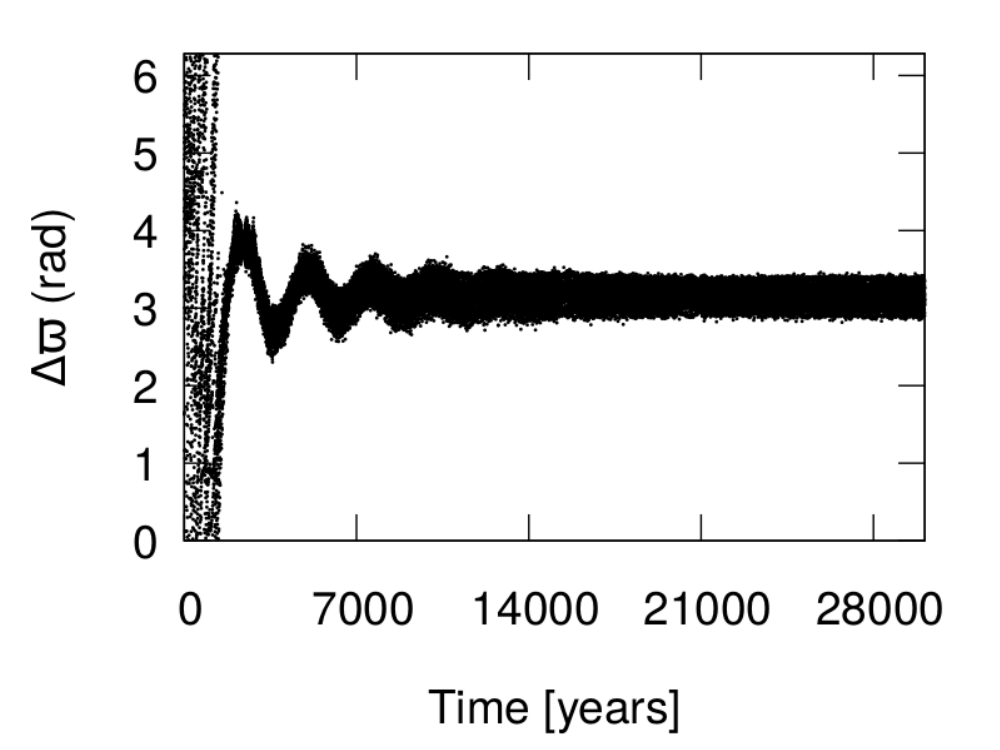}
\includegraphics[width=0.25\textwidth]{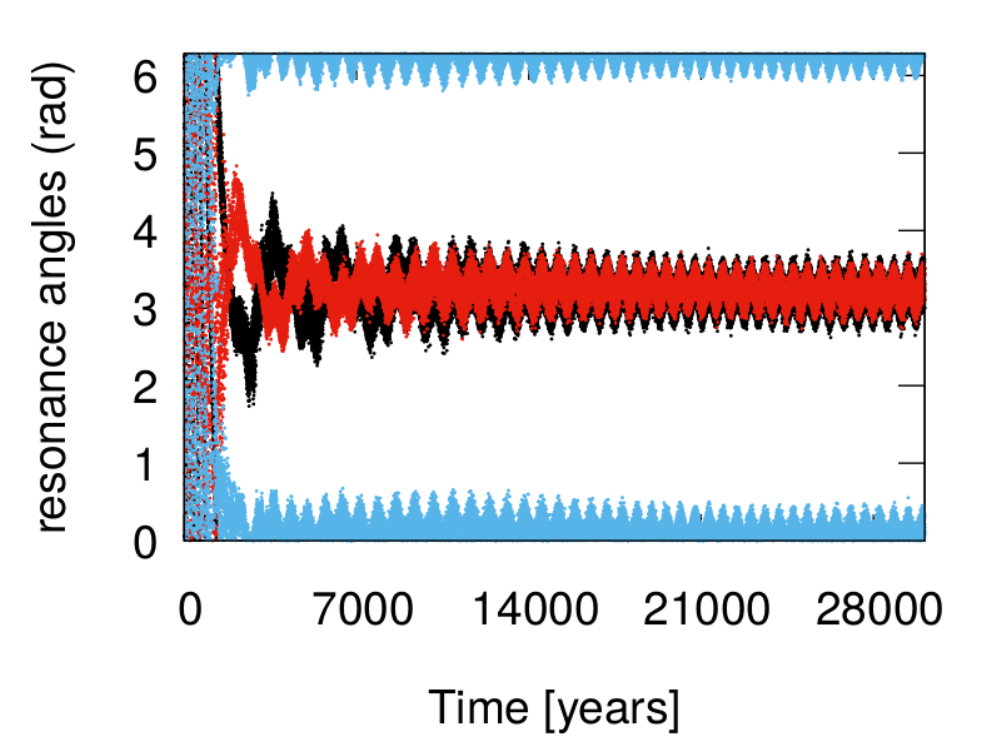}
}
}
}
\caption{Top panels: evolution of the period ratio (left), eccentricities $e_{1}$, $e_{2}$ (middle), and semi-major axes (right). The blue solid line in the top left panel shows the period ratio  9:7. Bottom panels: evolution of  $\Delta \varpi$ (left panel) and the resonance angles $\phi_{1}$ (black), $\phi_{2}$ (red), and  $\phi_{3}$ (blue) (right panel). The simulation parameters were $\Sigma_0 = 2 \times 10^{-5}$ and $\nu = 1 \times 10^{-6}.$ }
\label{fig:ecc10-example}
\end{figure*}

We illustrate the process of 9:7 resonance capture in a typical simulation contributing to the map for which disk parameters are $\Sigma_0 = 2 \times 10^{-5}$ and $\nu = 1 \times 10^{-6}$. The evolution of the period ratio, eccentricities $e_{1}$, $e_{2}$, $\Delta \varpi$ and the resonance angles $\phi_{1}$, $\phi_2$ and  $\phi_{3}$ are presented in Figure~\ref{fig:ecc10-example}. 

\begin{figure}[htb!]
\plotone{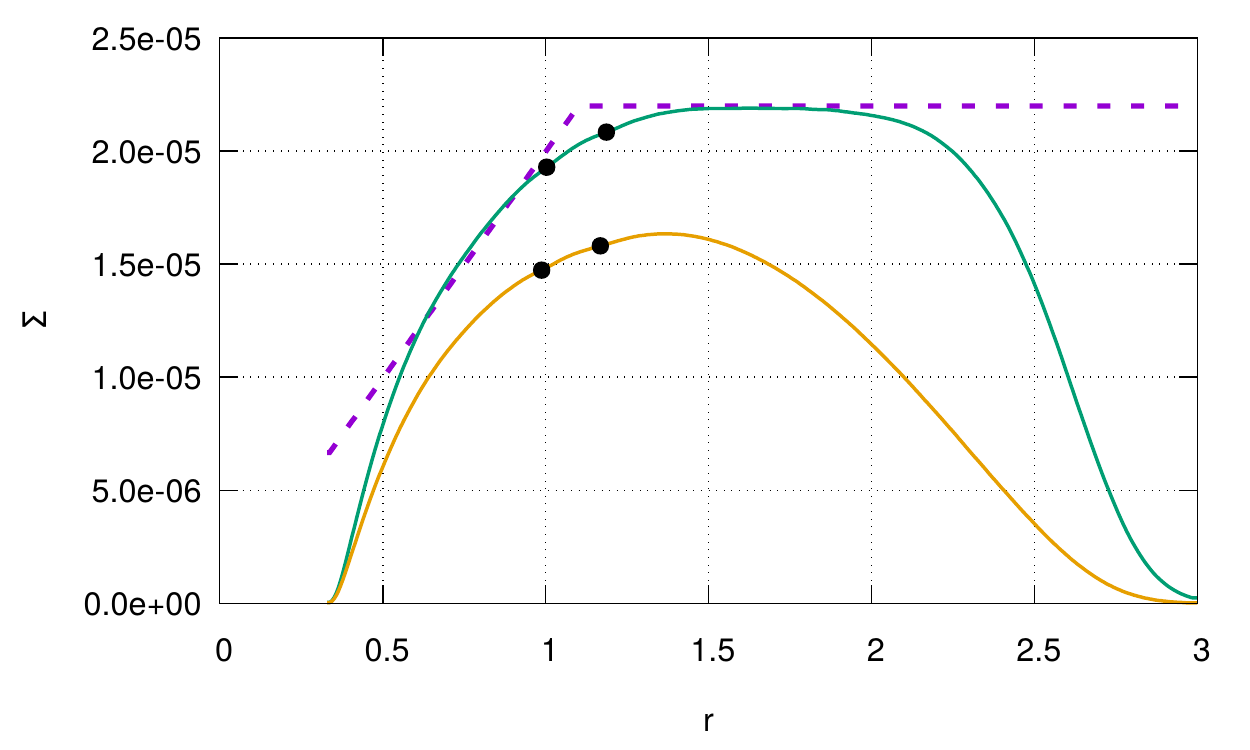}
\caption{Mean surface density profile of the disk at $t=0$ (dashed line), $t=1250$ yr (green solid line), and $t=15,000$ yrs (yellow solid line) in the simulation shown in Figure~\ref{fig:ecc10-example}. The positions of two planets are indicated by black filled circles.}
\label{fig:ecc10_sdensity}
\end{figure}

\begin{deluxetable}{ccccccc}
\tablecaption{The Parameters and Migration Times of Two Planets \label{tab:compare}}
\tablecolumns{7}
\tablenum{2}
\tablewidth{0pt}
\tablehead{
\colhead{Time} & \colhead{Planet} & \colhead{$\alpha$} & \colhead{$r_{p}$} & \colhead{$\Sigma_{p}$} & \colhead{$\tau_{\rm typeI}$} & \colhead{$\tau_{\rm fit}$} \\
\colhead{(yr)} & \colhead{} & \colhead{} & \colhead{} & \colhead{$(r_{p}^{2}/M_{\star})$} & \colhead{(Myr)} & \colhead{(Myr)}
}
\startdata
1250 & Inner & -0.55 & 1.003 & $1.929\times 10^{-5}$ & 0.89 & 0.90 \\
     & Outer & -0.31 & 1.187 & $2.084\times 10^{-5}$ & 0.94 & 0.64 \\
\hline
15,000 & Inner & -0.47 & 0.988 & $1.474 \times 10^{-5}$ & 1.09 & 1.02 \\
      & Outer & -0.29 & 1.168 & $1.581 \times 10^{-5}$ & 1.20 & 1.01 \\
\hline
30,000 & Inner & -0.40 & 0.975 & $9.530 \times 10^{-6}$ & 1.67 & 1.31 \\
      & Outer & -0.18 & 1.153 & $1.005 \times 10^{-5}$ & 1.76 & 1.68 \\
\hline
Averaged & Inner & ... & ... & ... & 1.22 & 1.08 \\
         & Outer & ... & ... & ... & 1.30  & 1.11 \\ 
\enddata
\end{deluxetable}

The result of this simulation is that after a short time of convergent migration with initial relative migration rate $\dot{a}/a \sim 0.96 ~{\rm Myr}^{-1}$, the two planets enter into the 9:7 resonance at $t \sim 1700$ yr and stay in this commensurability until the end of calculation. When the resonance capture occurs, $e_{2}$ is excited from zero to reach 0.004 and then evolves with $e_{1}$ showing similar behavior. 
In the final stages, eccentricities are oscillating around an equilibrium value ($\sim 0.004$) with $e_{2}$ slightly larger than $e_{1}$ while $\Delta \varpi$, $\phi_{1}$, and $\phi_{2}$ librate around $\pi$ and $\phi_{3}$ librates around zero. During the calculation, the two planets undergo slow migration without opening partial gaps in the disk. The evolution of the azimuthally averaged surface density, $\Sigma(r)$, is shown in Figure~\ref{fig:ecc10_sdensity}. The initial $\Sigma(r)$, which is defined through Equation~(\ref{disk}) is represented by the dashed line. We also plot  $\Sigma(r)$ at t = 1250 yrs (green solid line) and t = 15,000 yr (yellow solid line), thus showing the surface density profile before and after planets enter the resonance. The positions of the planets are indicated by black filled circles. Note that at the latest time the surface density has decreased significantly throughout the computational domain on account of viscous evolution and mass loss through the open boundaries, though this is not the case at the time of capture into the resonance.

It can be seen that planets cannot disturb the disk too much in this work, indicating that migration is in the type I regime. It is important to note that although the Lindblad torques that lead to inward migration are in the linear regime, this is not the case for corotation torques
that are expected to oppose them \citep[see][]{Paarde2009}. When unsaturated as expected here, these tend to be underestimated in a linear calculation, which nonetheless should provide migration rates that are correct to order of magnitude unless there is a near balance between corotation and Lindblad torques \citep[][]{Paarde2009}. According to \citet{Tanaka2002}, the timescale for type I migration $\tau=r_{p}/(-\dot{r_{p}})$ is given in the linear regime, by
\begin{equation}
\hspace{7mm} \tau=(2.7+1.1\alpha)^{-1}\frac{M_{\star}}{M_{p}}\frac{M_{\star}}{\Sigma_{p}r_{p}^{2}}\left(\frac{c_{s}}{r_{p}\Omega_{p}}\right)^2\Omega_{p}^{-1} \label{migration}
\end{equation}

\noindent where $\alpha$ is the slope of the disk defined through $\Sigma \propto r^{-\alpha}$, and $\Omega_{p}$ and $\Sigma_{p}$ are the angular velocity and the value of surface density at the location of a planet orbiting at distance $r_{p}$ from the central star.
In Table~\ref{tab:compare} we show $\alpha$, $r_{p}$ and $\Sigma_{p}$ for two planets, when $t=1250, 15,000$, and $30,000$ yr in the
simulation illustrated in Figure \ref{fig:ecc10_sdensity}, together with the type I migration time $\tau_{\rm typeI}$ for each planet calculated from Equation~(\ref{migration}). We also obtain and present in this table the migration time $\tau_{\rm fit}=a_{i}/(-\dot{a_{i}})$ ({\it i}=1,2 for inner and outer planet, respectively) fitted from the simulation. The fit has been done on 100 yr intervals around the moments of time mentioned in the table. 
Comparison of these values demonstrates that the migration of the planets is reasonably close to that found from Equation (\ref{migration}), indicating that indeed the type I migration regime applies. It can be convincingly seen from the last row of the table, where the averaged values of migration time are given. 

A semianalytic solution for two planets in a second-order resonance undergoing migration and orbital circularization is described in \citet{Xiang2015}. After the system is locked in the resonance, a relationship between the eccentricities $e_1,$ $e_2,$ the circularization time, $\tau_{c,i}=-e_{i}/\dot{e_{i}}$, and the migration time, $\tau_{\rm mig,i}=-2a_{i}/\dot{a_{i}}$, for planet $i$ with $i= 1,2$ is obtained, which takes the form
\begin{eqnarray} 
\hspace{-12mm}&& \frac{1}{m_1m_2}\left(\frac {m_1 a_2 e_1^2 }
{\tau_{c,1} (1-e_1^2)} + \frac {m_2a_1 e_2^2 }{\tau_{c,2} (1-e_2^2)}\right)=\nonumber\\
\hspace{-12mm}&&\frac{a_2^{3/2}\sqrt{(1-e_2^2)}-a_1^{3/2}\sqrt{(1-e_1^2)}}
{m_2\sqrt{a_2(1-e_2^2)}+m_1\sqrt{a_1(1-e_1^2)}}
\left( \frac{1}{\tau_{\rm mig,2}}-\frac{1}{\tau_{\rm mig,1}}\right)
\label{ejcons}
\end{eqnarray}
\noindent Since $\phi_{1}$ and $\phi_{3}$ are librating, the rate of change of $\phi_{3}-\phi_{1}$ is neglected. An expression that can be used to
determine the eccentricity ratio of two planets is given as 
\begin{eqnarray}
\hspace{-10mm}&&\sum_{n,i }\hspace{-1mm}
\left[\frac{\sqrt{1-e^2_1}m_2}{n_1a^2_1a_2e_1}\frac{\partial F_{i}^{n}}{\partial e_1}
 -\frac{\sqrt{1-e^2_2}m_1}{n_2a^3_2e_2}\frac{\partial F_{i}^{n}}{\partial e_2}\right]\hspace{-1mm}\cos\psi_{i}^{n}=0
\label{eratio} \end{eqnarray}
\noindent where $\psi_{i}^{n}\equiv~(2n-i)\phi_{3}~-~(n-i)\phi_{1}$. $F_{i}^{n}$ are the amplitudes of the direct parts of the disturbing function, $R_{D},$ appropriate for second-order resonances, $n$ is a non-negative integer such that $0  \le   n \le 2$,
and $i$ is a positive or negative integer or zero such that $ -1 \le i \le 4$. The form of $F_{i}^{n}$ in which terms up to fourth order in the eccentricities can be included, as indicated in \citet{Xiang2015}, can be found in \citet{Murray1999}. 

In order to compare with the analytic model, first we measure the values of $e_{1}$, $e_{2}$, together with $\tau_{c}$ and $\tau_{\rm mig}$ for each planet in the simulation illustrated in Figure \ref{fig:ecc10_sdensity} to check whether they satisfy the relationship given by Equation~(\ref{ejcons}). 
After $t = 11,000 $ yr, when the planets are captured into 9:7 resonance, we obtain $a_{1} = 0.992$, $a_{2}=1.17308$, $e_{1}=0.00265$, and $e_{2}=0.00326$ from the calculation.
The circularization time and migration time of each planet are fitted from the data in the interval $t = 11000 \pm 100 $ yr and the results are $\tau_{c,1}=4.70 \times 10^{3} $ yr, $\tau_{c,2}=3.39 \times 10^{3} $ yr, $\tau_{\rm mig,1}=1.75 \times 10^{6} $ yr and $\tau_{\rm mig,2}=1.64 \times 10^{6} $ yr. Inserting these numbers into the left-hand side  of Equation~(\ref{ejcons}) we obtain $5.41 \times 10^{-4}$ while the right-hand side yields $5.70 \times 10^{-4}$, which shows that the relation between those parameters in the simulation is consistent with the model mentioned above.

We go on to compare the ratio of the eccentricities of the two planets in the simulation with the value predicted from Equation~(\ref{eratio}). 
If we only consider the last stages of the simulation, $\phi_{1}$ librates around $\pi$ and $\phi_{3}$ librates around zero. Since both resonant angles librate with small amplitude, we take $\phi_{1}$ to be $\pi$ and $\phi_{3}$ to be zero, and then $\psi_{i}^{n}=-(n-i)\pi$. Because the final values of $e_{1}$ and $e_{2}$ are very close to zero, terms that include $e_{1}$ or $e_{2}$ at an order higher than two in Equation~(\ref{eratio}) can be neglected. In this way, for a system locked in a 9:7 second-order resonance with very small eccentricities, the eccentricity ratio is found to depend only on the planet mass ratio. Considering our case in which $m_{1}=m_{2}$, we obtain $e_{1}/e_{2}=0.96$, in good agreement with the result of the simulation.

\begin{figure}[htb!]
\includegraphics[width=\columnwidth]{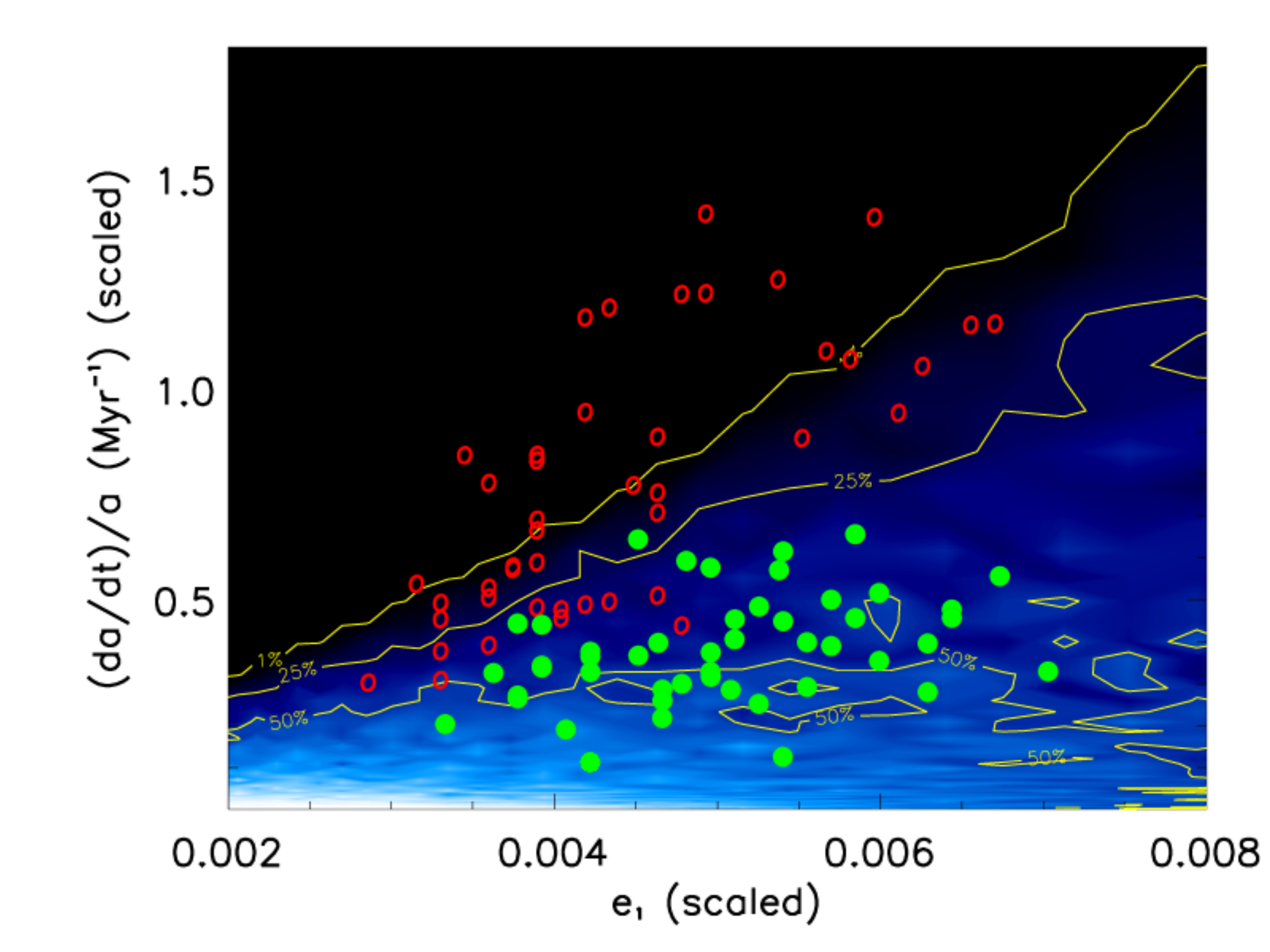}
\caption{Results of simulations with initial $e_{1}=0.005$ and $e_{2}=0$ presented in the (rescaled eccentricity, rescaled relative migration rate) plane. The red open circles and green filled circles represent the ``fail'' and ``capture'' cases, respectively. The contour plot from \citet{Mustill2011} shows the capture probability of 9:7 resonance in a system of a test particle and planet of mass of $3M_{\oplus}$. The contour lines for $1\%$, $25\%$, and $50\%$ are indicated by yellow lines.}
\label{fig:ecc10_comparison}
\end{figure}

\citet {Mustill2011} give a capture probability associated with a second-order resonance depending on rescaled eccentricity and relative migration rates in a restricted three-body problem. For a system consisting of an inner test particle and an outer planet, the eccentricity of the inner test particle $e_{1}$ and the relative migration rate $\dot{a}/a$ are rescaled to dimensionless quantities according to
\begin{equation}
 J_{1}=k\left(\frac{m_{2}}{M_{\oplus}}\right)^{-1}\left(\frac{M_{\star}}{M_{\odot}}\right)e_{1}^{2} \quad {\rm and}
 \label{eq:dimensionless1}
\end{equation}

\begin{equation}
\dot{\beta}=l\left(\frac{m_{2}}{M_{\oplus}}\right)^{-2}\left(\frac{M_{\star}}{M_{\odot}}\right)^{3/2}\left(\frac{a_{1}}{\rm 1au}\right)
\left(\frac{a_{2}}{\rm 1au}\right)^{1/2}\left(\frac{\dot{a}/a}~{\rm Myr^{-1}}\right),
\label{eq:dimensionless2}
\end{equation}

{\noindent}where $k$ and $l$ are 125,717 and 143.877, respectively, for 9:7 resonance capture. Adopting $m_{2}=3M_{\oplus}$ and $M_{\star}=M_{\odot}$ in Equations~(\ref{eq:dimensionless1}) and (\ref{eq:dimensionless2}), we can obtain their estimate of the probability of capture into 9:7 resonance as a function of $e_{1}$ and $\dot{a}/a$ for a system consisting of an inner test particle and an outer planet of mass of $3M_{\oplus}$ in a circular orbit. As well as for the restricted three-body problem case, the rescaled parameters for an unrestricted two-planet system are also given in the Appendix of their paper. These are given as
\begin{equation}
 J_{1}'= \frac{3[(2-j)^{2}+(m_{1}/m_{2})\alpha_{0}^{2}j^{2}]}{32(m_{2}/M_{\star})\alpha_{0}f_{45}}e_{1}^{2}
 \label{eq:dimensionless3}
\end{equation}

\begin{equation}
\dot{\beta}'=\frac{3(j-2)a_{1}^{1/2}(\dot{a}/a)}{16(m_{2}/M_{\star})^{2}\alpha_{0}^{2}f_{45}^{2}},
\label{eq:dimensionless4}
\end{equation}
{\noindent}where $j=9$, $\alpha_{0}=(7/9)^{2/3}$ and $f_{45}=14.3866$ for 9:7 resonance. However, the capture probability for this case is not given explicitly in their paper.

In order to make a comparison between results of our hydrodynamic simulations (for $m_{1}=m_{2}=3M_{\oplus}$, the initial eccentricities $e_{1}=0.005$ and $e_{2}=0$) and the results of \citet {Mustill2011}, we first calculate $J_{1}'$ and $\dot{\beta}'$ for each simulation using the  values of the eccentricities and the relative migration rate at the point of entry into resonance. We then obtain new values of $e_{1}$ and $\dot{a}/a$ from Equations~(\ref{eq:dimensionless1}) and ~(\ref{eq:dimensionless2}) after assuming the equivalences $J_{1}'\equiv J_{1}$ and $\dot{\beta}' \equiv\dot{\beta}$. In this way the eccentricity $e_{1}$ and relative migration rate $\dot{a}/a$ for a system with two planets, as in our simulations, are rescaled to enable a comparison with a system consisting of a planet and test particle.

We enter the results corresponding to the rescaled $e_{1}$ and $\dot{a}/a$ obtained from our simulations onto the contour plot showing the capture probability into 9:7 resonance provided by \citet {Mustill2011} assuming an outer planet mass is equal to $3M_{\oplus}$ and a central solar mass in a restricted three-body problem. The result is illustrated in Figure~\ref{fig:ecc10_comparison}. The regions with capture probability of $50\%$, $25\%$, and $1\%$ are indicated by yellow contour lines. The results of hydrodynamic simulations are represented by circles. The red open circles and green filled circles indicate ``fail'' and ``capture'' cases, respectively. In this figure, we can see that all ``capture'' cases are inside of the region with capture probability larger than $1\%$, and all but one of them are inside the region with capture probability larger than $25\%.$
The planets pass through the resonance when the rescaled relative migration rate is higher than 0.7 ${\rm Myr}^{-1}$. In the region with low relative migration rate (rescaled $\dot{a}/a < 0.7 ~{\rm Myr}^{-1}$), planets are captured more easily if they have higher eccentricity. The results of our hydrodynamic simulations are thus consistent with the contour plot of \citet {Mustill2011}.

Recently, capture conditions for second-order resonances in a system of two comparable low-mass planets have been presented in \citet{Xu2017}.
According to their analysis, which is based on a restricted three-body model, in order to capture the planets into a second-order resonance, the migration timescale $T_{m}'$, the eccentricity damping timescale $T_{e}'$ and the eccentricity of the inner planet when entering the resonance $e_{0}$ must satisfy
\begin{equation}
T_{e}' \gtrsim \frac{P_{1}}{8\pi\mu}, ~ ~ T_{m}' \gtrsim \frac{P_{1}}{8\pi\mu^{2}}{\rm ln}\frac{\mu}{e_{0}^{2}}, ~ ~ e_{0} \lesssim \mu^{1/2}
\label{eq:xu}
\end{equation}
where $T_{e}'=T_{e,1}=-e_{1}/\dot{e_{1}}$ and $T_{m}'=1/(1/T_{m,2}-1/T_{m,1})$  with $T_{m,i}= -(\dot{a_{i}}/a_{i}+2e_{i}^{2}/T_{e,i})^{-1}$. Here $T_{e,i}$ is the circularization time for planet $i,$ with $i=1,2$ denoting the inner and outer planet, respectively, and $\mu=(m_{1}+m_{2})/M_{\star}$. Considering the parameters of our simulation, we obtain the conditions for resonance capture based on Equation~(\ref{eq:xu}) as $T_{e} \gtrsim 2.2\times 10^{3} $ yr, $T_{m} \gtrsim 3.3\times 10^{8} $ yr and $e_{0} \lesssim 0.0042$. Although, in agreement with our findings, Equation (\ref{eq:xu}) formally predicts that for specified disk parameters the conditions for capture are not satisfied for sufficiently small eccentricity (see below), from Figure~\ref{fig:ecc10_contour_refit}, we can see that the relative migration rates in our survey are much higher than the values mentioned above. Accordingly, capture is not expected to be certain.

In the regime considered here, where orbital circularization is included, a second-order resonance trapping needs nonzero initial eccentricities. For this reason, it is interesting to determine what is the minimum initial value of the eccentricity for which the capture can take place in our simulations. We have performed the calculations with $e_{2}=0$ and varying the value of $e_{1}$. In results, the minimum initial value of $e_{1}$ for which we obtained a capture was 0.002. We have also tried $e_{1}=0.001$, but in this case the system passes through the 9:7 resonance because  $e_{1}$ is damped to a value close to zero before arriving there. In other words, the initial eccentricity did not survive the journey to resonance from outside against circularization. 


\subsection{Dependence on the Initial Orbit of the Outer Planet} \label{sec:initial-orbit}

\begin{deluxetable}{ccc}
\tablecaption{The Disk Parameters of Selected Simulations \label{tab:shift}}
\tablecolumns{3}
\tablenum{3}
\tablewidth{0pt}
\tablehead{
\colhead{Case} & \colhead{$\Sigma_{0}$} & \colhead{$\nu$}
}
\startdata
1 & $4 \times 10^{-5}$ & $3 \times 10^{-6}$ \\
2 & $5.5 \times 10^{-5}$ & $4 \times 10^{-6}$ \\
3 & $4.5 \times 10^{-5}$ & $5 \times 10^{-6}$ \\
\enddata
\end{deluxetable}

\begin{figure}[htb!]
\plotone{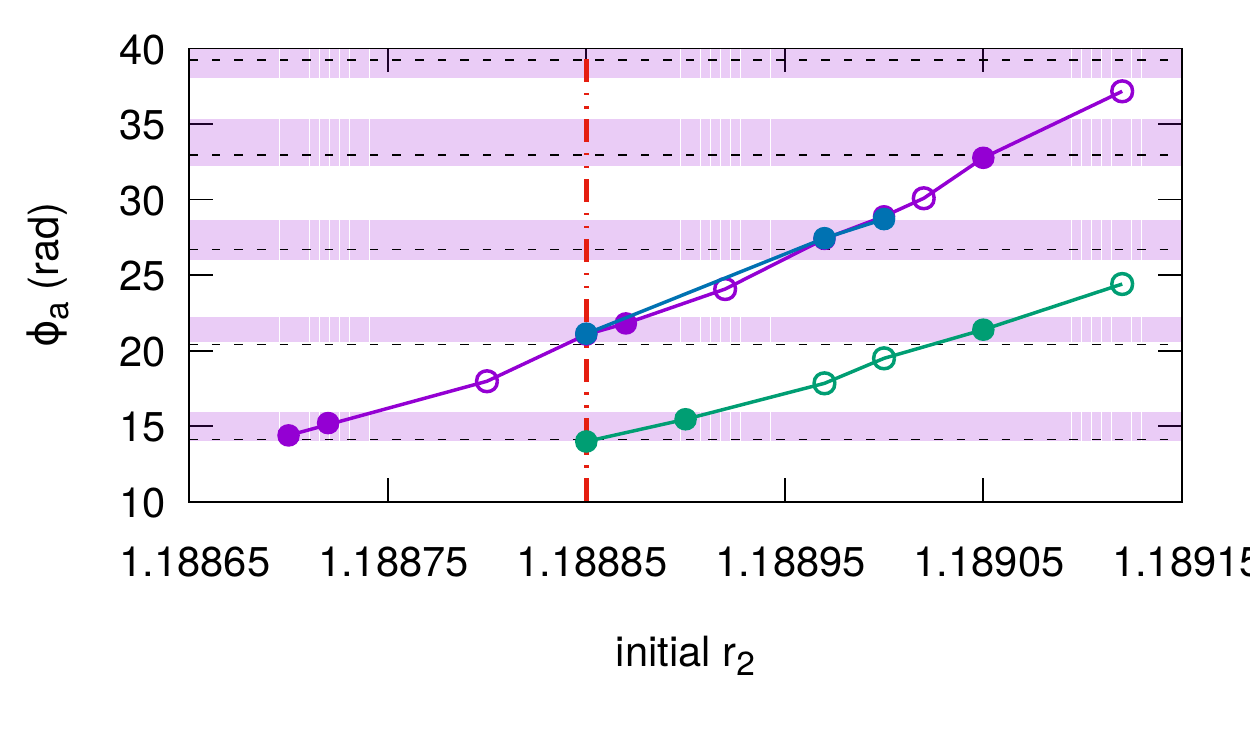}
\caption{Changes of the accumulated resonant angle $\phi_{a}$ with initial radial locations of the outer planet $r_{2}$ for the three simulations chosen from Figure \ref{fig:ecc10_map} (see text for more details).
The filled circles represent ``capture'' cases, and the open circles represent ``fail'' cases. Simulations indicated by different colors have different initial disk parameters.
Violet corresponds to $\Sigma_0= 4 \times 10^{-5}$ and $\nu = 3 \times 10 ^{-6}$, green to $\Sigma_0 = 5.5 \times 10^{-5}$ and $\nu=4 \times 10^{-6}$ and blue to  $\Sigma_0 = 4.5 \times 10^{-5}$ and $\nu = 5 \times 10^{-6}$. The violet strips are the regions of capture as shown in Figure~\ref{fig:ecc10_map}. The red dot-dashed vertical line indicates the initial value of $r_{2}=1.18885$ as adopted in the simulations shown in Figure \ref{fig:ecc10_map}.}
\label{fig:ecc10_ini}
\end{figure}

As we mentioned above, all the previous simulations start from a planet configuration in which the initial orbital radius of inner planet $r_{1}$ is equal to 1 and that of outer planet $r_{2}$ is equal to 1.18885. In order to remove the limitation of fixed starting configuration and obtain a more general picture of 9:7 resonance capture, we show the results of simulations starting with the same disk and planet parameters as previously but different initial $r_{2}$. From the simulations used to construct the map, we choose three for which planets are captured in 9:7 resonance. The disk parameters for them are given in Table~\ref{tab:shift}. 
For each case we recalculate the evolution of the planets starting from various initial values of $r_{2}$ and obtain new results, which are illustrated in Figure~\ref{fig:ecc10_ini}. In order to maintain the relative migration rate and circularization rate similar to the values in the previous simulations, we take the initial value of $r_{2}$ to be in the small interval [1.18865, 1.18915].

The disk parameters used in simulations are represented by different colors. The red dashed line indicates the initial $r_{2}$ adopted in the original map simulations. Thus, the three points on this line show the results of three cases from the map simulations (the violet and blue points on the red dot-dashed line overlap each other). We also indicate the regions of $\phi_{a}$ for capture by violet stripes, which are identical to those in Figure~\ref{fig:ecc10_map}.

Although the very small changes in the initial value of  $r_{2}$ in the new simulations ensure similar values of the relative migration rates and circularization rates to those for the previous cases, the values of $\phi_{\rm entry}$ and $N_{\rm rot}$ are significantly changed, making the accumulated resonance angles different from those  obtained in the map simulations. It can be seen from Figure~\ref{fig:ecc10_ini} that for simulations with the same disk parameters, starting from a larger $r_{2}$, the planets enter into 9:7 resonance with higher values  of $\phi_{a}$ and vice versa. 
Thus, results could be affected by the sensitivity of $\phi_{a}$ to a particular choice of the initial value of $r_{2}$ adopted in the simulation. However, we can still predict whether the planets will be captured or pass through the 9:7 resonance by reference of the new value of $\phi_{a}$ to the regions for capture for the map simulations. If the accumulated resonance angle remains in the same region for capture or moves to another violet stripe region on account of starting from another value of $r_{2}$, the planets can be locked in 9:7 resonance.
On the other hand, when $\phi_{a}$ is shifted outside of those regions as a result of changing $r_{2}$, the planets will pass through the resonance. From the new results from case 1 (violet) and case 3 (blue) we note that the relation between $\phi_{a}$ and $r_{2}$ is very similar. Based on this, we predict that for other simulations with the same accumulated resonance angle, in the map simulations, but with different disk parameters, the results should have the same dependence on $r_{2}$.


\subsection{Trajectories in the ($e_1{\rm cos}\phi_{1}$,~$e_1{\rm sin}\phi_{1}$) Plane} \label{sec:track-plane}

\begin{figure*}[htb!]
\plottwo{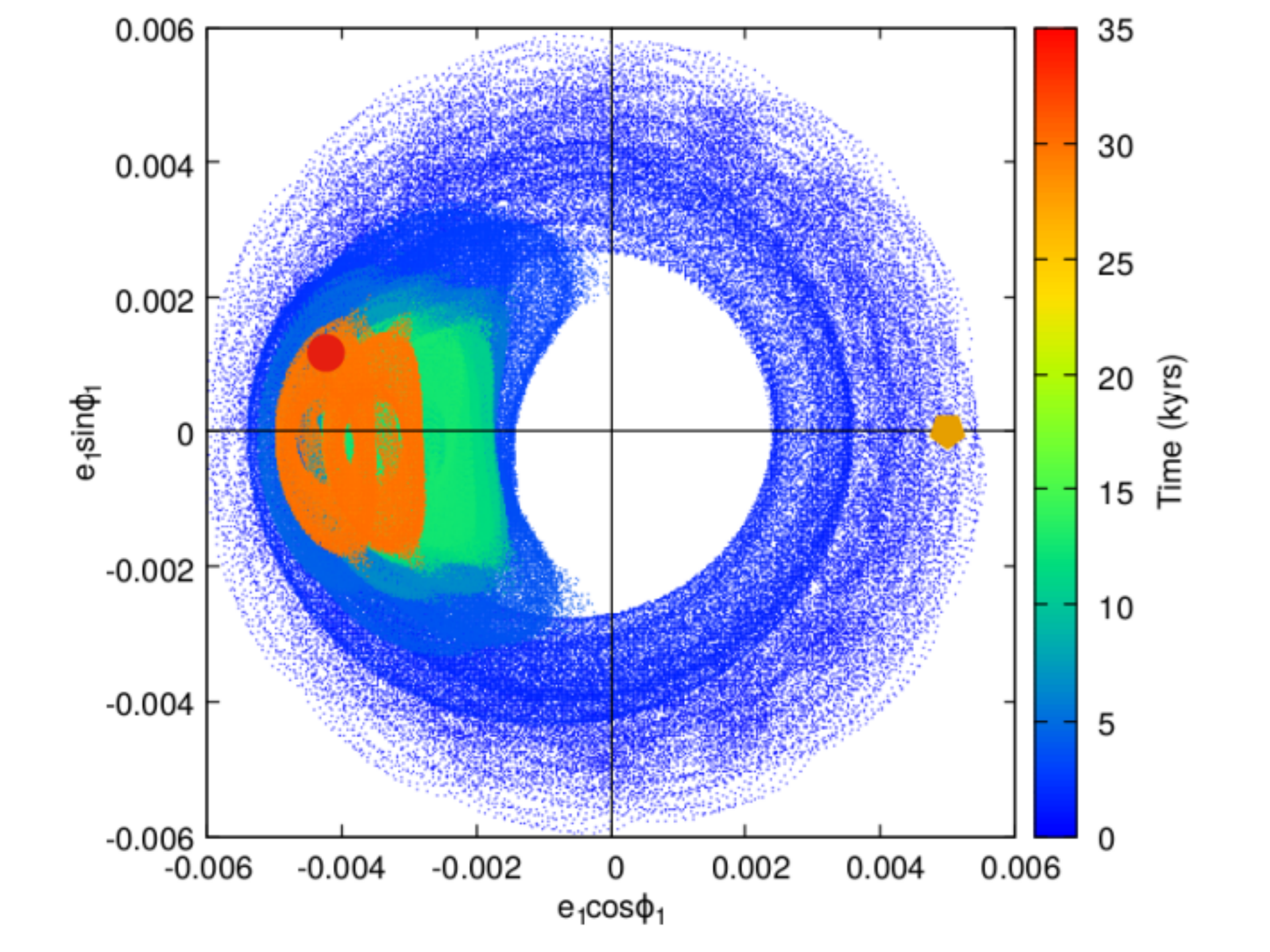}{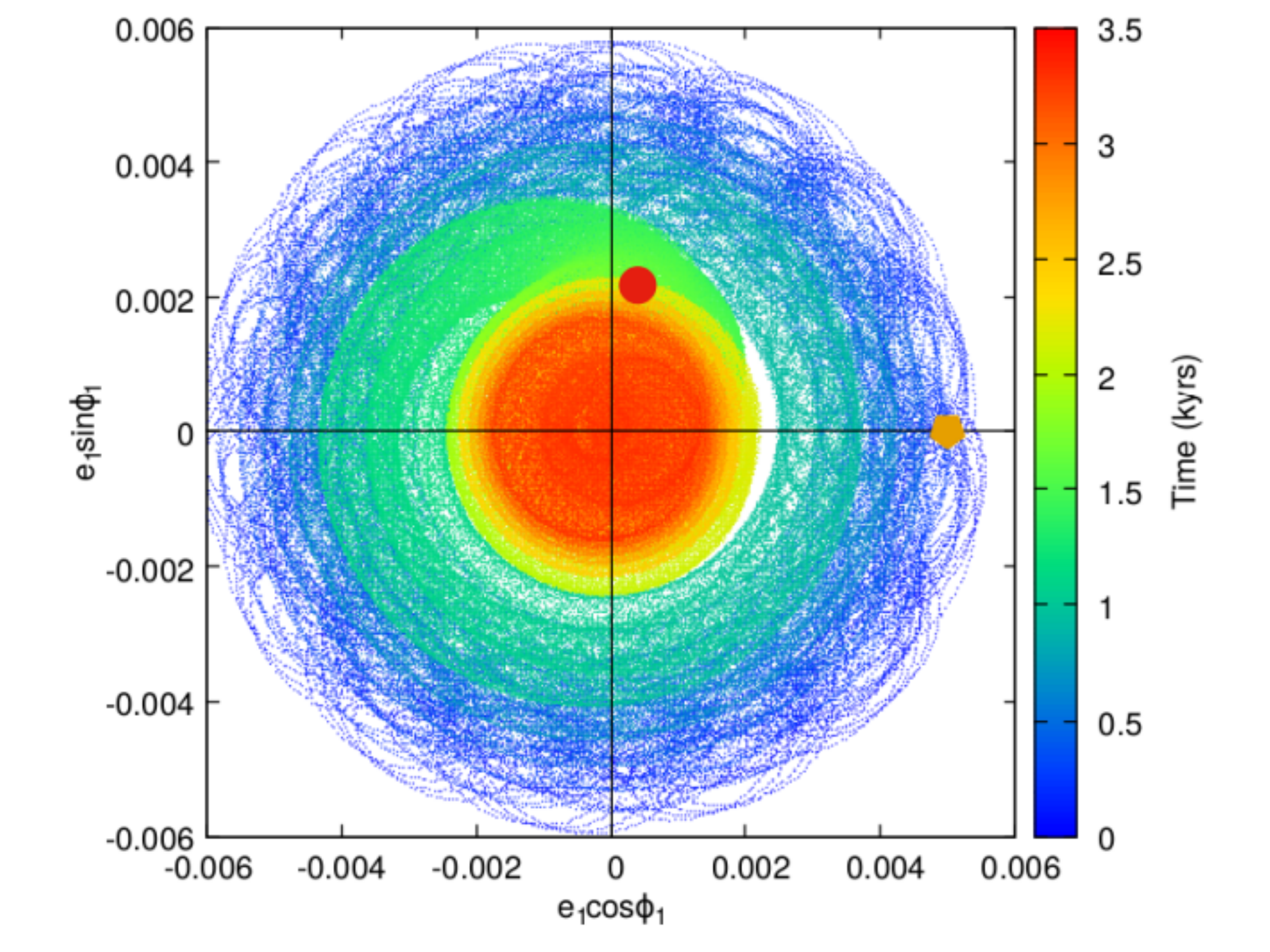}
\caption{Trajectories in the ($e_{1}{\rm cos}\phi_{1}$, $e_{1}{\rm sin}\phi_{1}$) plane during migration for two cases shown in the map presented in Figure \ref{fig:ecc10_map}. The left panel is for $\Sigma_0 =2 \times 10^{-5}$, $\nu = 1 \times 10^{-6}$ and the right panel is for $\Sigma_0 =3 \times 10^{-5}$, $\nu = 4 \times 10^{-6}$. The orange pentagons show the initial positions of the system, and the red circles show their position when the systems arrive at the point where $\phi_{a}$ is specified (see text).}
\label{fig:plane}
\end{figure*}

\begin{figure}[htb!]
\includegraphics[width=\columnwidth]{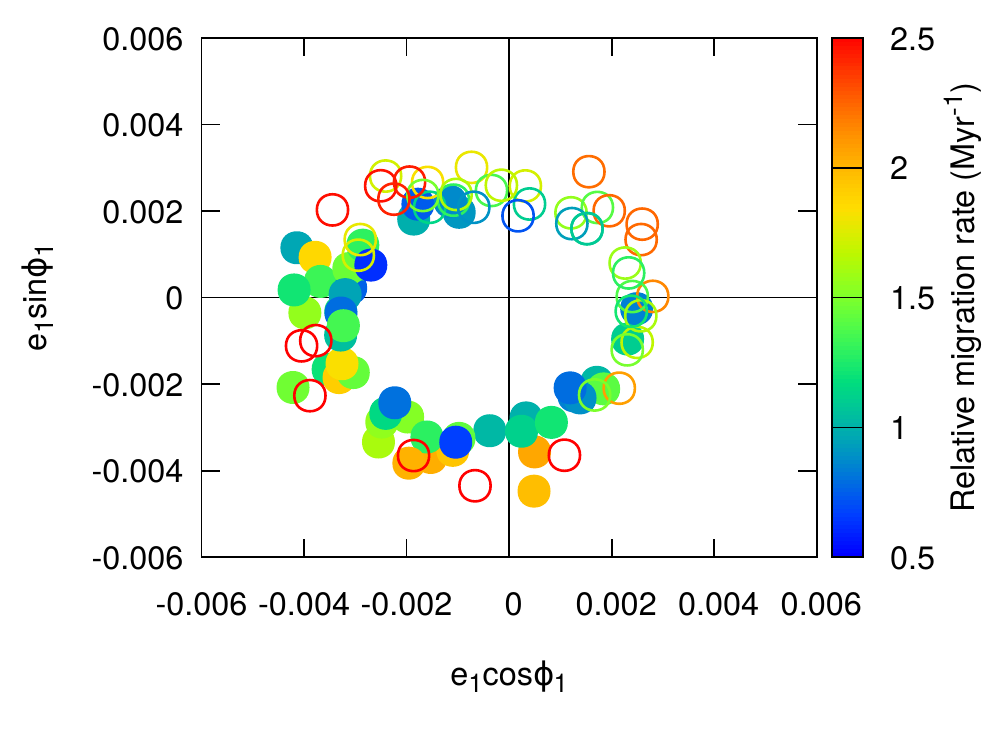}
\caption{Entry locations in the ($e_{1}{\rm cos}\phi_{1}$, $e_{1}{\rm sin}\phi_{1}$) plane for the runs shown in the map presented in Figure \ref{fig:ecc10_map} with relative migration rate smaller than $2.5 ~{\rm Myr}^{-1}$. The filled and open circles indicate the ``capture'' cases and the ``fail'' cases, respectively (see text for more details).}
    \label{fig:ecc10_plane}
\end{figure}

For the restricted three-body problem, the particle's Hamiltonian curve can be drawn in the ($\sqrt{2\Phi}{\rm cos}\phi$,$\sqrt{2\Phi}{\rm sin}\phi$) plane, where $\sqrt{2\Phi}$ is a scaled eccentricity of the particle and $\phi$ is its resonance angle (\citet{Murray1999}). Here we consider the case of  a 9:7 resonance capture for two equal-mass planets. In order to describe the motion of the inner planet undergoing capture into a 9:7 resonance during migration, we plot the trajectory in the  ($e_{1}{\rm cos}\phi_{1}$, $e_{1}{\rm sin}\phi_{1}$) plane for two cases in Figure~\ref{fig:plane}. 

The left panel shows the track of the inner planet in a ``capture'' case with $\Sigma_0 = 2 \times 10^{-5}$ and $\nu = 1 \times 10^{-6}$ while a ``fail'' case is presented in the right panel for which $\Sigma_0 = 3 \times 10^{-5}$ and $\nu = 4 \times 10^{-6}$. Given the initial $e_{1} = 0.005$ and $\phi_{1} = 0$, the track of the inner planet in two cases starts from $(0.005, 0)$, indicated by an orange pentagon on the plane.
Initially the trajectory rotates clockwise as a result of the circulation of $\phi_{1}$. However, $e_{1}$ decreases, resulting in the radial distance of the curve from the origin to becoming smaller. The location on the plane when the planets enter into the resonance is denoted by a red circle on the plot. After arriving at the resonance, the trajectories of the two cases illustrated behave differently. In the ``fail'' case, the planets pass through the resonance. The resonance angle $\phi_{1}$ circulates while $e_{1}$ decreases. In this way the track contracts toward the origin. In the ``capture'' case shown in the left hand panel, $\phi_{1}$ begins to librate while $e_{1}$ increases. The trajectory passes through, $e_{1}{\rm sin}\phi_{1}=0,$ while $e_{1}{\rm cos}\phi_{1} < 0,$ which indicates that $\phi_{1}$ is librating around $\pi.$ In addition, $e_{1}$ oscillates with large amplitude. 
     
Comparing plots of this kind, we see that the system can enter into the resonance from different locations on the plane. In Figure~\ref{fig:ecc10_plane}, we show the entry locations in the calculations presented in Figure \ref{fig:ecc10_map}. It is easy to notice that the first quadrant, which covers the angles in the range of $[0, \frac{1}{2}\pi]$, is occupied only by the ``fail'' cases (open circles) independently of the relative migration rate. In the ``capture'' cases, the faster-migrating planets (orange filled circles) tend to enter the resonance in the third quadrant, which covers the angles in the range of $[\pi, \frac{3}{2}\pi]$ while the entries of the slower-migrating planets (blue and green filled circles) are widely distributed in the second, third, and fourth quadrants. Instead, in the ``fail'' cases for the high migration rates, the red open circles are located in all quadrants. The above discussion implies that there is an entrance on the plane for locking the planets into 9:7 MMR and that the existence, width, and the location of the entrance should depend on physical parameters such as the relative migration rate and the circularization rate, which can also be seen in Figure \ref{fig:dot_angle}. Only if the planets follow trajectories that find this entrance can they be captured.


\subsection{Capture in 9:7 Resonance with High Initial Eccentricities} \label{sec:ecc}

\begin{figure*}[htb!]
\centerline{
\vbox{
\hbox{
\includegraphics[width=0.25\textwidth]{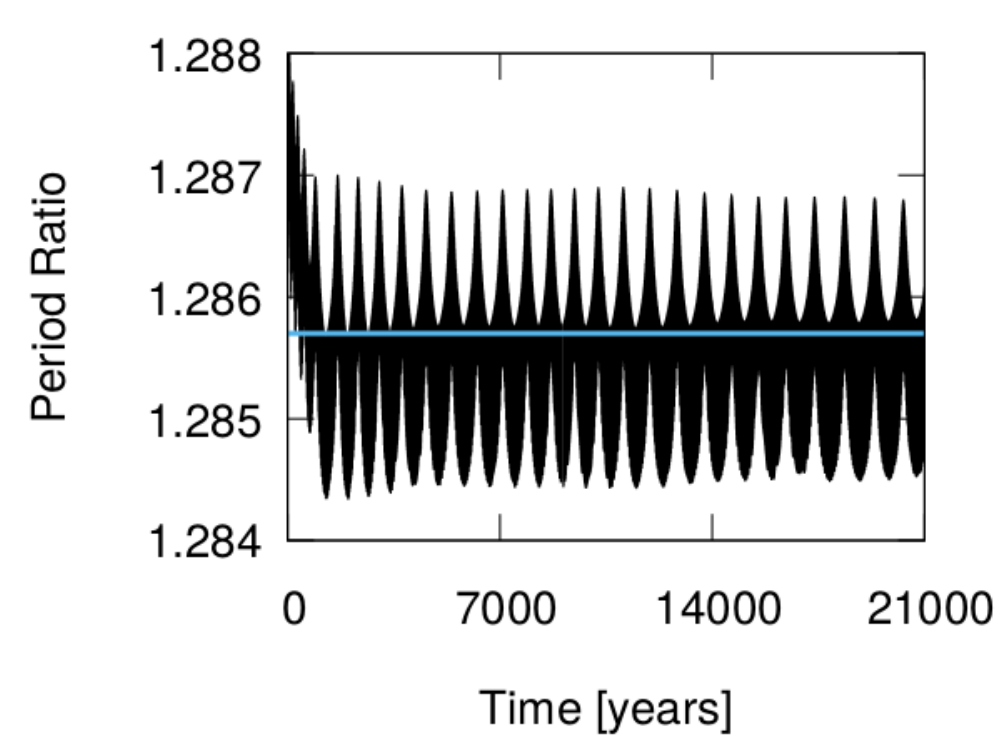}
\includegraphics[width=0.25\textwidth]{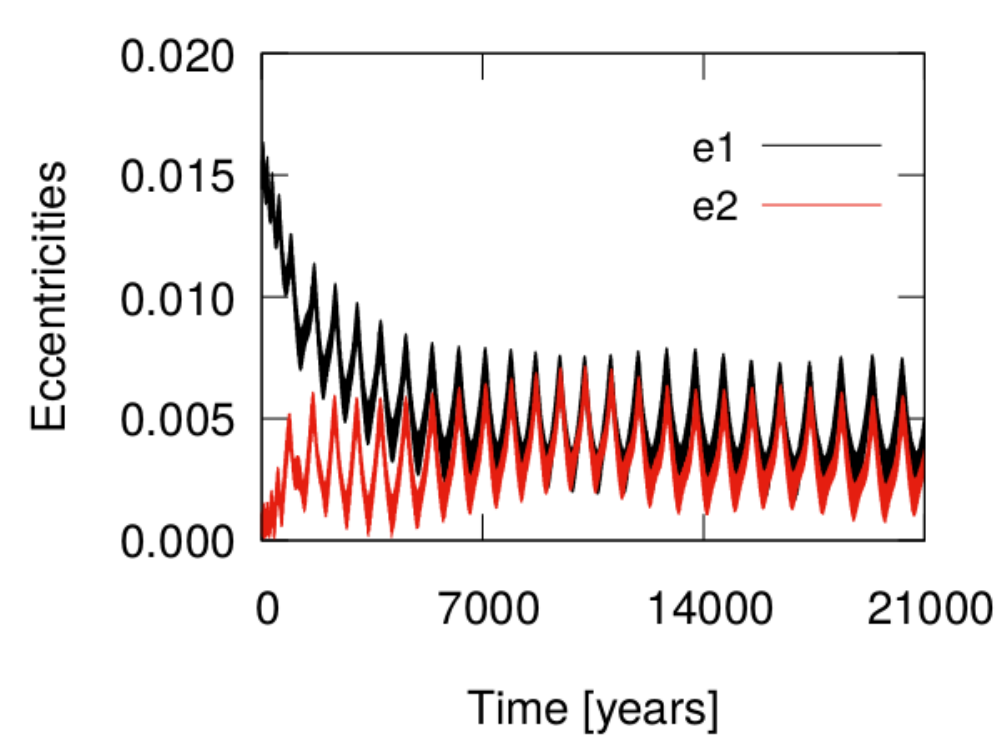}
\includegraphics[width=0.25\textwidth]{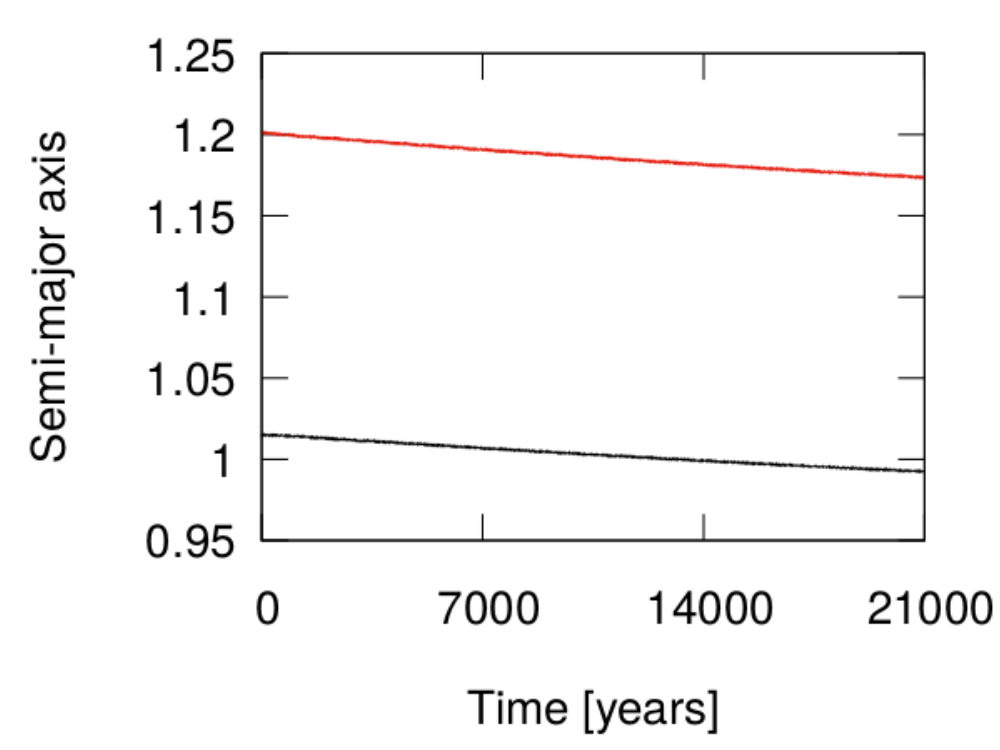}
}
\hbox{
\includegraphics[width=0.25\textwidth]{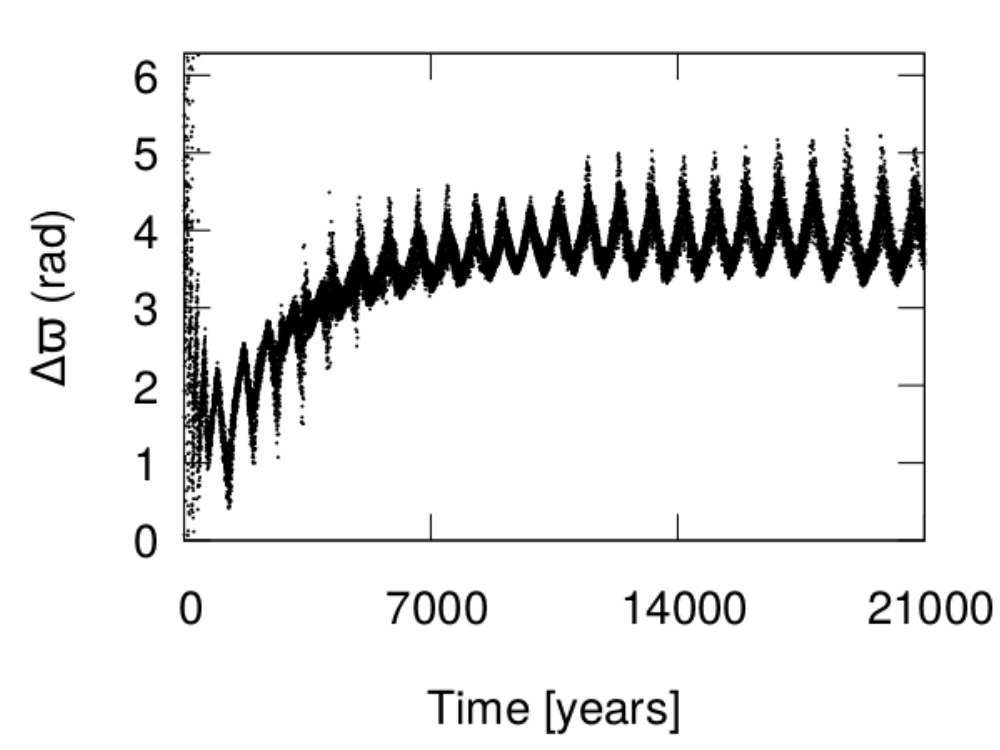}
\includegraphics[width=0.25\textwidth]{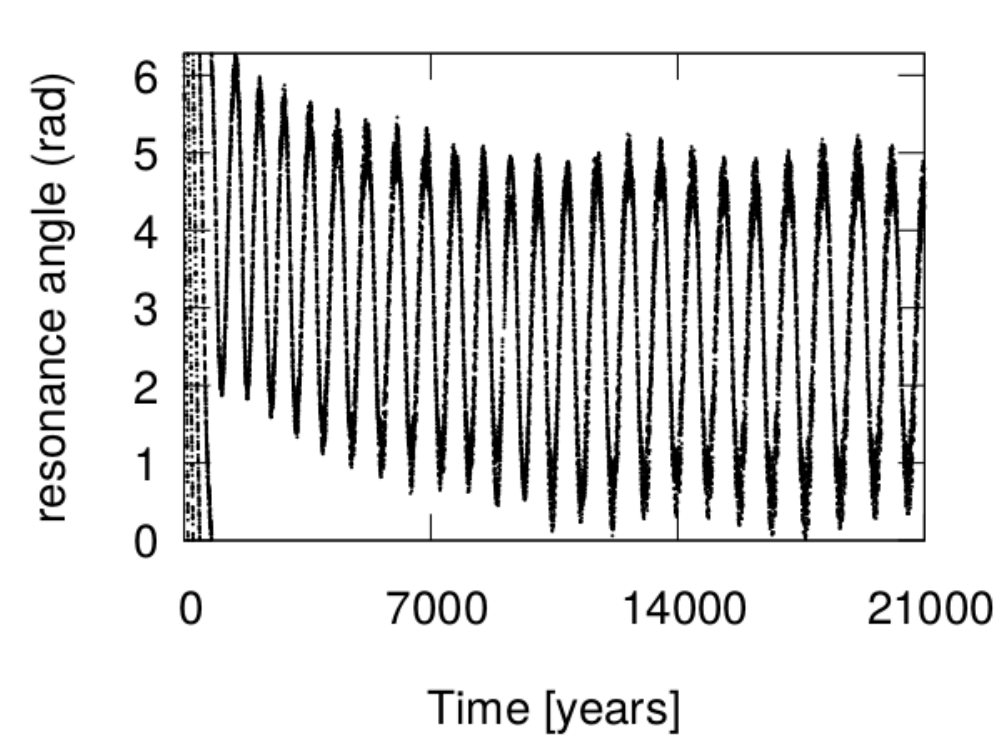}
\includegraphics[width=0.25\textwidth]{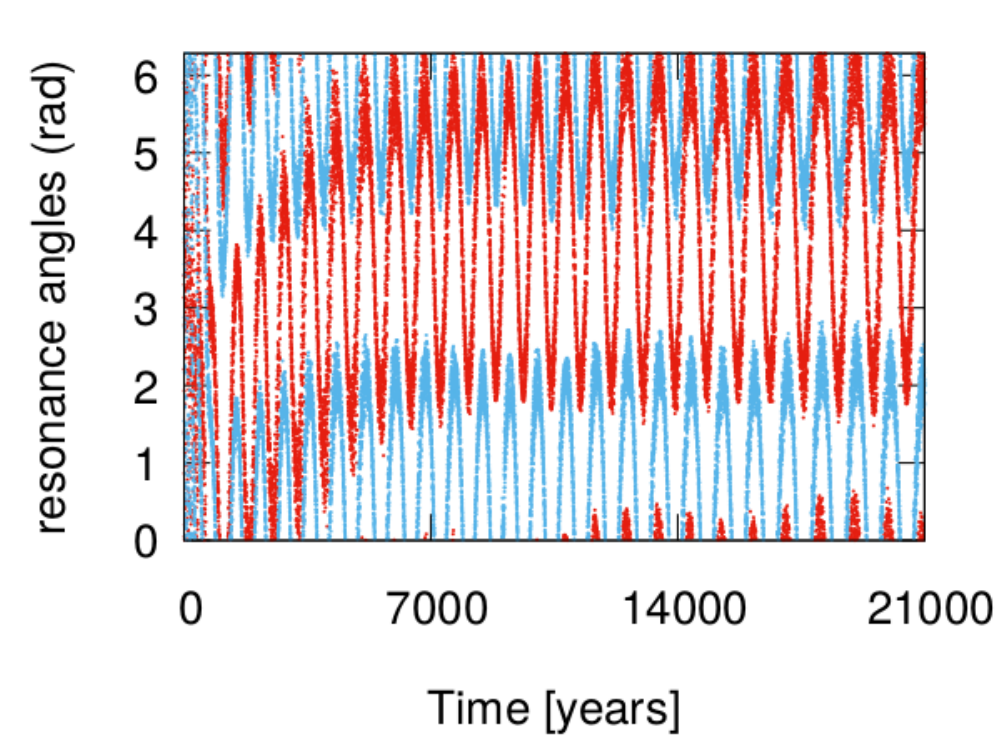}
}
}
}
\caption{Top panels: evolution of the period ratio (left), eccentricities $e_{1}$, $e_{2}$ (middle), and the semi-major axes (right) for the simulation with $\Sigma_0 = 2 \times 10^{-5}$ and $\nu = 1 \times 10^{-6}$. The horizontal solid line in the top left panel shows the period ratio 9:7. Bottom panels: from left to right, $\Delta \varpi$, the resonance angle $\phi_{1}$, and the resonant angle $\phi_{2}$ (red) together with the resonant angle $\phi_3$ (blue). The initial eccentricities for this case were $e_{1}= 0.015$ and $e_{2}=0.$ }
\label{fig:ecc9-example}
\end{figure*}

In previous sections, we presented the results of hydrodynamic simulations modeling equal-mass planets in a protoplanetary disk with initial $e_{1}=0.005$ and $e_{2}=0$. We found the regions of $\phi_{a}$ associated with 9:7 resonance capture in simulations with different disk parameters. The width of those regions depends on the relative migration rate and circularization rate. However, the issue of what happens for planets with significantly higher initial eccentricities remains to be addressed. In order to investigate this, we present here the results of simulations with various disk parameters but for which the planets have higher initial eccentricities.

First, we consider 9:7 resonance capture in a system of two equal-mass planets ($m_{1}=m_{2}=3M_{\oplus}$) with initial eccentricities $e_{1}=0.015$ and $e_{2}=0$. One example is given in Figure~\ref{fig:ecc9-example} for which $\Sigma_0=2 \times 10^{-5}$ and $\nu = 1 \times 10^{-6}$.
After a short period of convergent migration, the planets arrive at the 9:7 resonance at $t \sim 1400$ yr with relative migration rate $\dot{a}/a \sim 0.79 ~{\rm Myr}^{-1}$. Then, they are captured into the resonance as $e_{2}$ is excited. At the end of calculation, $e_{1}$ and $e_{2}$ are oscillating around 0.004. 
$\Delta \varpi$ is found to librate in the range  [3.2, 5.1]. The resonance angles $\phi_{1}$ and $\phi_{2}$ librate around 2.8 and 4, while $\phi_{3}$ librates around 0.2. Compared with the results shown in Figure~\ref{fig:ecc10-example}, the final values of eccentricities are seen to be similar, but the centers of libration of the resonance angles and $\Delta \varpi$ are shifted by modest amounts from zero or $\pi$. It can be also noticed that the period ratio, eccentricities, and resonance angles oscillate with larger amplitudes in this simulation.

Based on this example, we present a series of simulations in which the disk parameters are chosen such that $\Sigma_0$ ranges between $4.5 \times 10^{-5}$ and $6 \times 10^{-5}$ and $\nu$ lies in the range of $1 \times 10^{-6} - 7 \times 10^{-6}$. The results are shown in Figure~\ref{fig:ecc9_region} as a function of the disk parameters and $\phi_{a}$. The value of $\phi_{a}$ in this series of simulations is measured when the semi-major axis ratio reaches 1.1827. This number has been obtained from Equation (\ref{X22}) for $e_{1}=0.012$, which corresponds to the eccentricity of the inner planet when the planets arrive at 9:7 resonance.
The filled and open circles represent the ``capture'' and ``fail'' cases, respectively. The distribution of $\phi_{a}$ also shows several particular regions for capture indicated by red stripes in the figure. Among them, the lowermost region for capture is around $\phi_{\rm entry}=2\pi$ which is represented by a dashed line. In addition to this one, there are another three narrow red stripes in the upper part of the figure. It is noted that the planets can be captured in 9:7 resonance with initial $e_{1}=0.015$ but pass through the resonance with initial $e_{1}=0.005$ for the disk parameter space adopted in this series of simulations.  
Moreover, the regions of $\phi_{a}$ for capture in this parameter space are divided into several narrow components in the interval  $[2\pi , 4\pi]$. In contrast, we find only one continuous region for capture in any interval of $2\pi$ in the previous series of simulations.

\begin{figure}[htb!]
\includegraphics[width=\columnwidth]{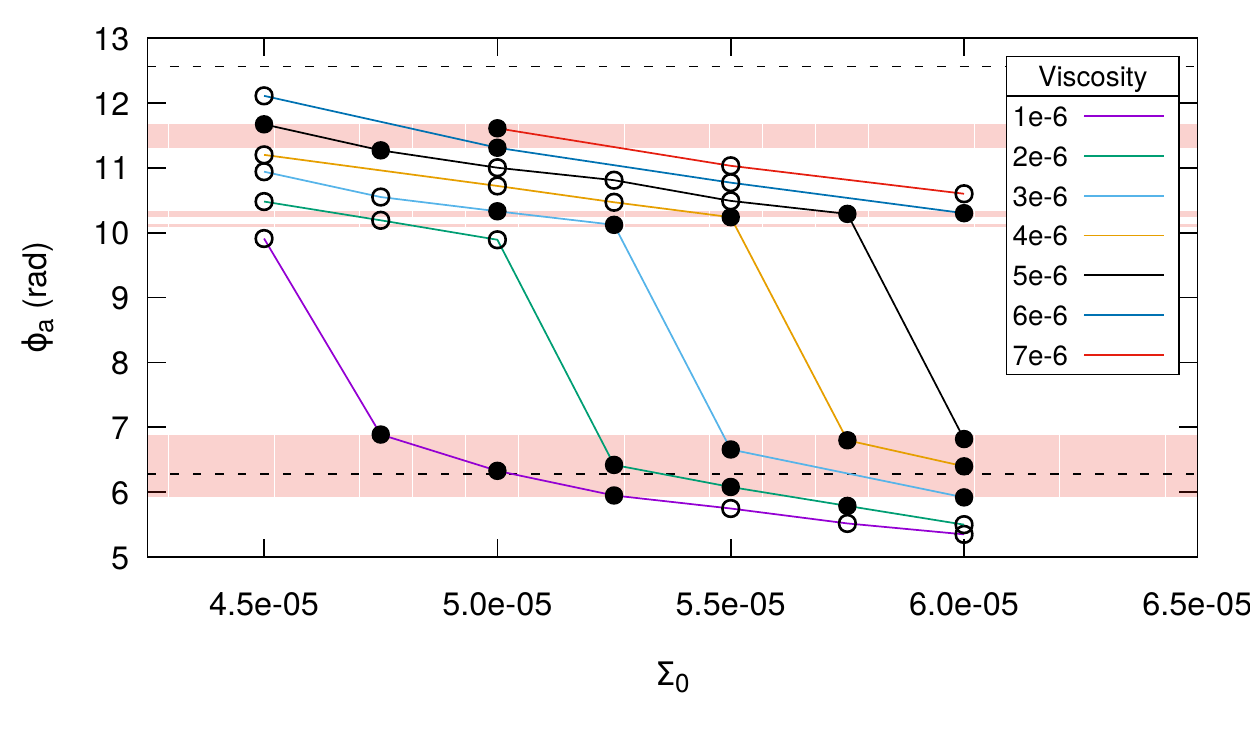}
\caption{Accumulated resonance angle $\phi_{a}$ as a function of surface density scaling parameter, $\Sigma_0,$ and viscosity, $\nu,$ 
from the results of simulations with initial $e_{1}=0.015$ and $e_2=0$. The filled and open circles indicate ``capture'' cases and ``fail'' cases, respectively. The dashed line indicates the location of $\phi_{a}$ for which $\phi_{\rm entry}=2\pi$. Red regions represent the ranges of $\phi_{a}$  where capture takes place.}
\label{fig:ecc9_region}
\end{figure}

In order to describe the motion of the planet in a calculation where there is capture into  9:7 resonance, we draw the track of the inner planet 
in the ($e_{1}{\rm cos}\phi_{1}$, $e_{1}{\rm sin}\phi_{1}$) plane during migration. We consider a ``capture'' case with $\Sigma_0=5 \times 10^{-5}$ and $\nu = 3 \times 10^{-6}$ in Figure~\ref{fig:ecc9_plane}. The track rotates clockwise before arriving at the resonance. At the same time, $e_{1}$ decreases, which results in a reducing radius of the trajectory in the plane. In the final stages, the track of the planet is confined to a small region colored dark orange, and the system is locked in the 9:7 resonance. The location where $\phi_{a}$ is specified for the 9:7 commensurability is also indicated in the plane. The entrance for capture into resonance is very narrow. 

\begin{figure}[htb!]
\plotone{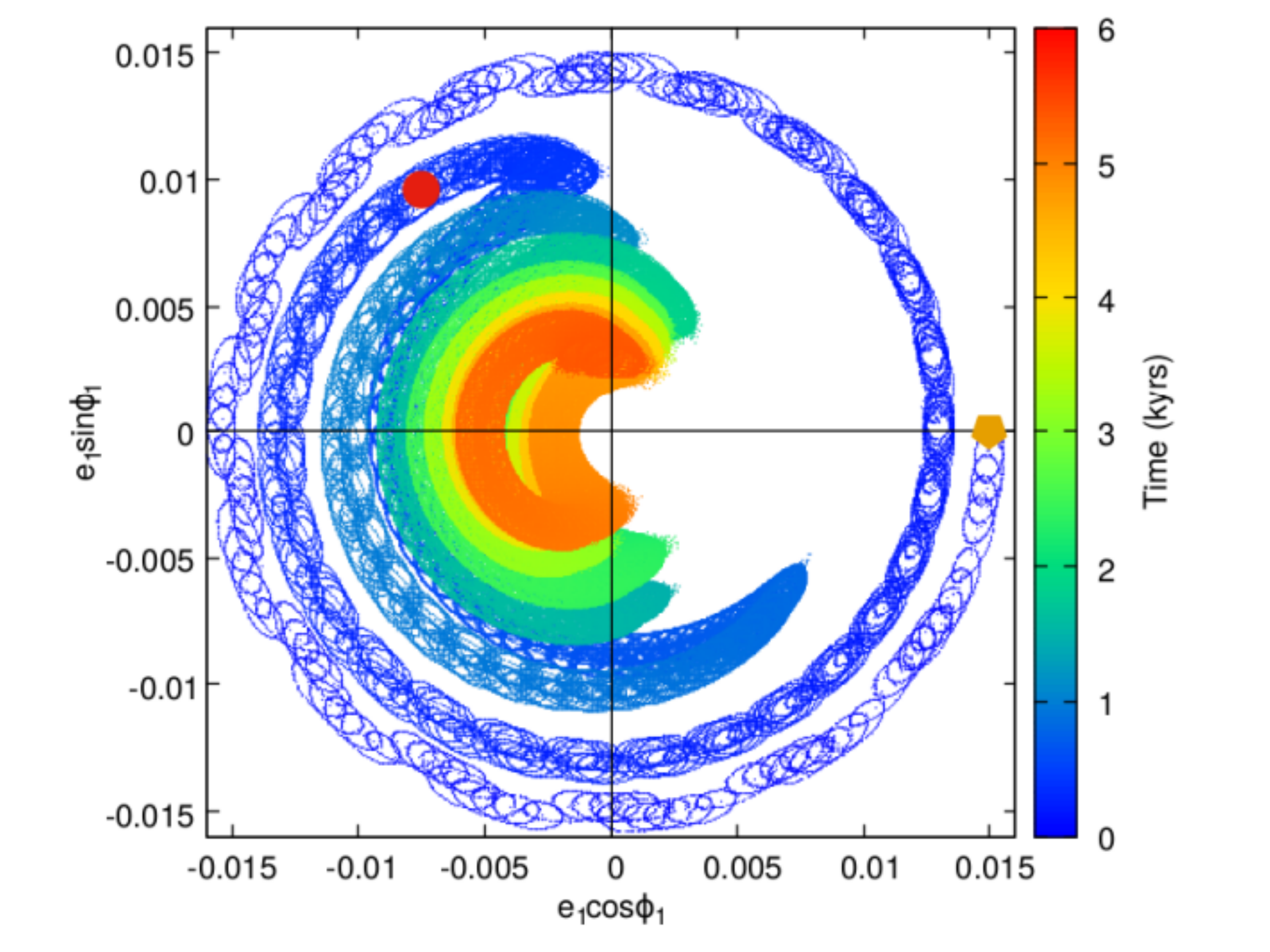}
\caption{Trajectory of the system in the ($e_{1}{\rm cos}\phi_{1}$, $e_{1}{\rm sin}\phi_{1}$) plane for a case with initial $e_{1}=0.015$ and $e_2=0$. The simulation was for $\Sigma_{0}=5 \times 10^{-5}$, $\nu = 3 \times 10^{-6}$. The symbols have the same meaning as in Figure~\ref{fig:plane}.}
\label{fig:ecc9_plane}
\end{figure}

\begin{figure*}[htb!]
\centerline{
\vbox{
\hbox{
\includegraphics[width=0.25\textwidth]{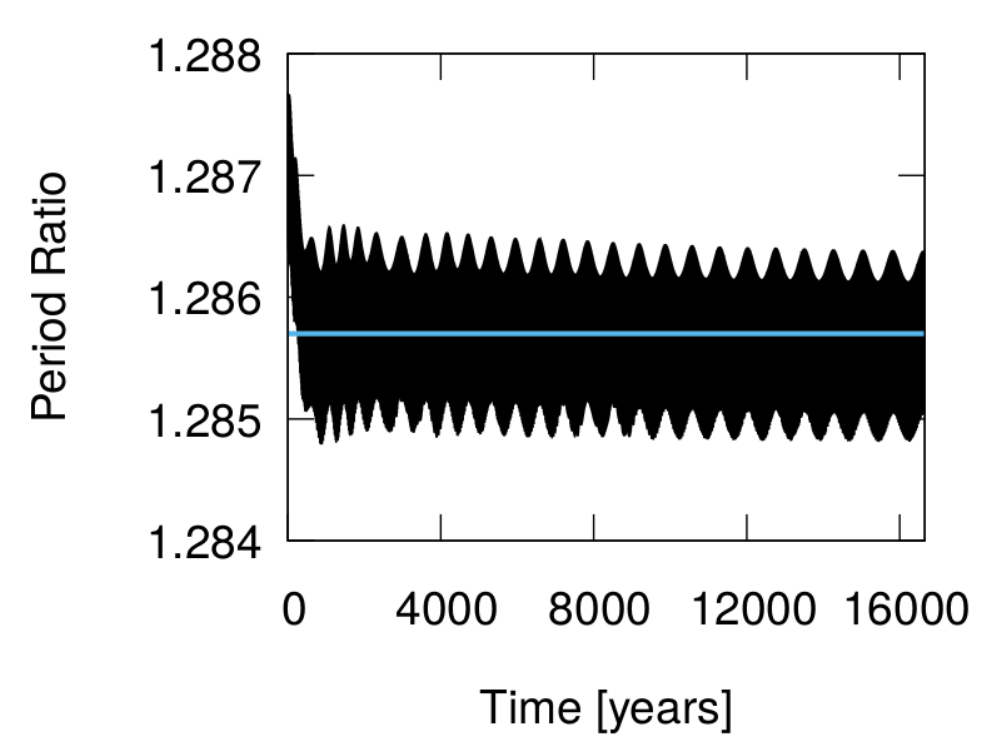}
\includegraphics[width=0.25\textwidth]{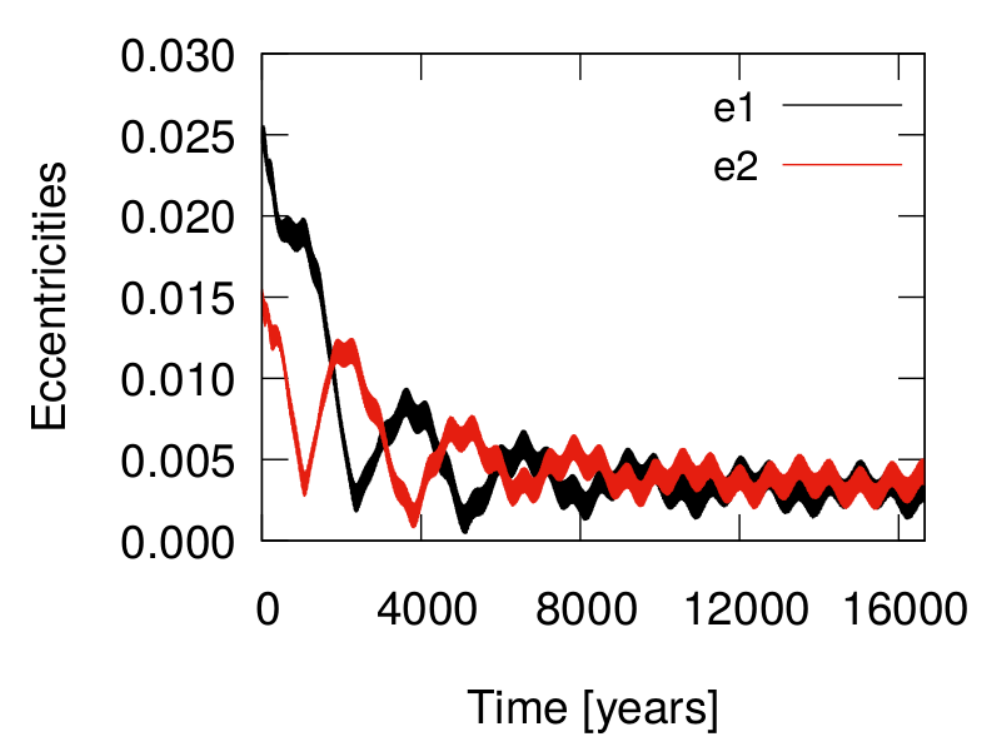}
\includegraphics[width=0.25\textwidth]{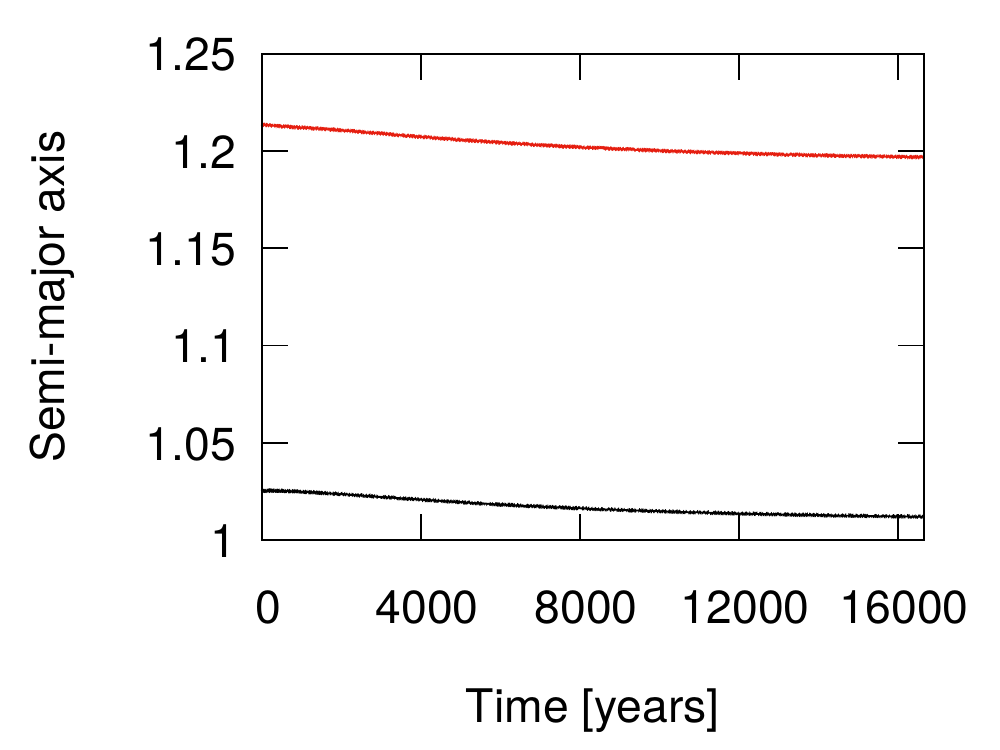}
}
\hbox{
\includegraphics[width=0.25\textwidth]{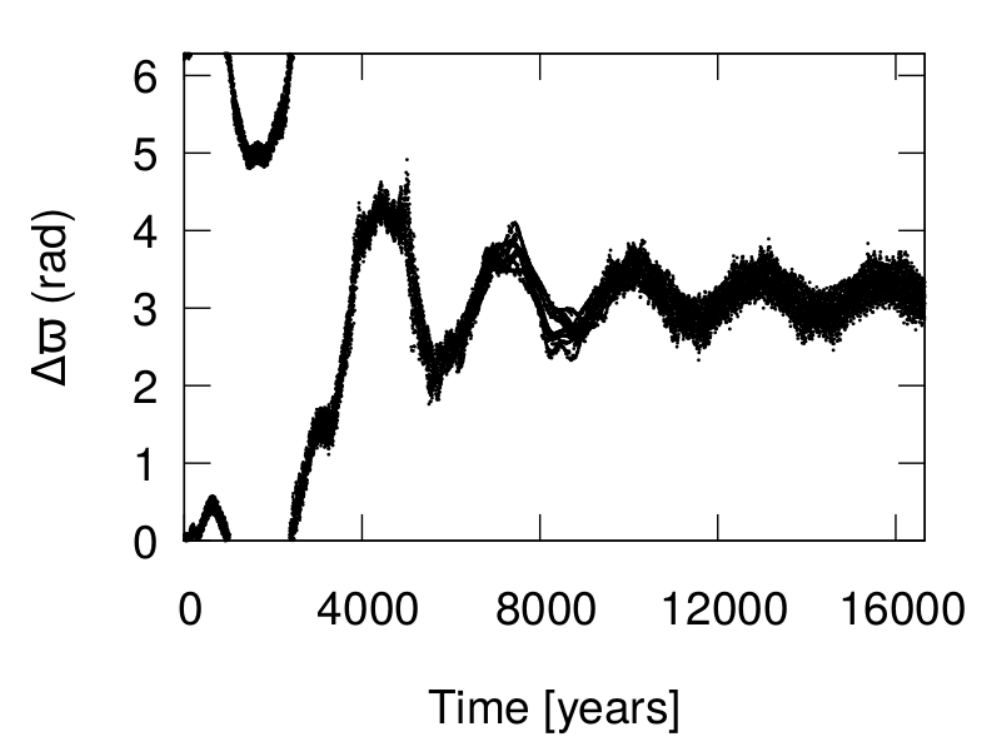}
\includegraphics[width=0.25\textwidth]{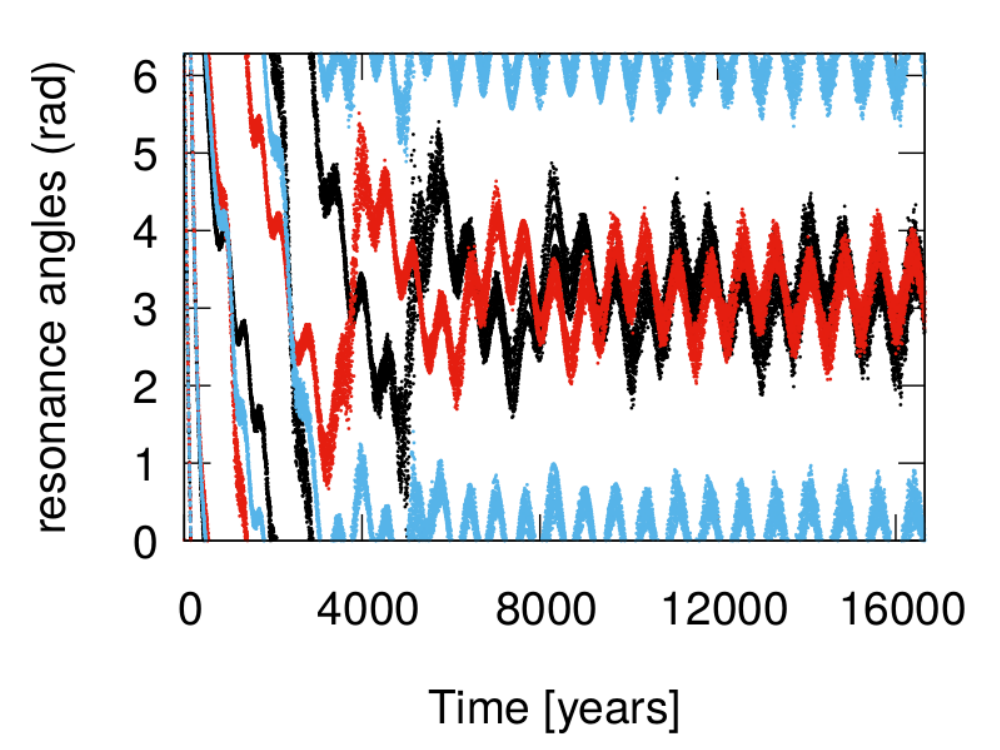}
}
}
}
\caption{Top panels: evolution of the period ratio (left), eccentricities $e_{1}$, $e_{2}$ (middle), and the semi-major axes (right) for the simulation with $\Sigma_0 = 3 \times 10^{-5}$ and $\nu = 5 \times 10^{-6}$. The horizontal solid line in the top left panel shows the period ratio 9:7. Bottom panels: from left to right, $\Delta \varpi$ and the resonance angles $\phi_{1}$ (black), $\phi_{2}$ (red), and $\phi_3$ (blue). The initial eccentricities for this case were $e_{1}= 0.025$ and $e_{2}=0.015.$ }
\label{fig:move26}
\end{figure*}

In addition, we consider a system of two equal-mass planets ($m_{1}=m_{2}=3M_{\oplus}$) with initial $e_{1}=0.025$ and $e_{2}=0.015$. The disk parameters were $\Sigma_0 = 3 \times 10^{-5}$ and $\nu = 5 \times 10^{-6}$. In this simulation, both eccentricities of two planets are 
initially higher than in previous cases. The evolution is illustrated in Figure~\ref{fig:move26}. The eccentricities $e_{1}$ and $e_{2}$ initially damp very rapidly during convergent migration. Initially, $\Delta \varpi$ oscillates around zero and all the resonant angles circulate until $t \sim 600$ yr when the period ratio is approaching 9:7. Subsequently, the period ratio of the two planets maintains in this commensurability. 
The general trend of the evolution is that the eccentricities decrease with large-amplitude oscillations present until $t \sim 6000$ yr. 
During this period, the center of oscillation of $\Delta \varpi$ moves from zero to $\pi$ and the behavior of the resonant angles transforms from circulation to libration. After $t \sim 6000$ yr, $e_{1}$ and $e_{2}$ tend to equilibrium values with small amplitude of oscillations superposed. All the resonant angles librate with amplitudes that decrease with time. At the end of the evolution $e_{1}$ and $e_{2}$ are around $0.004$ and $\Delta \varpi$ librates around $\pi$. The final values of all parameters are similar to the results of the simulation with initial  $e_{1}=0.005$ and $e_{2}=0$.

Based on the above example, we show the results of a series of simulations for which $\Sigma_0$ is chosen in the range from $1.5 \times 10^{-5}$ to $4 \times 10^{-5}$ and $\nu$ in the range from $1 \times 10^{-6}$ to $8 \times 10^{-6}$ in Figure~\ref{fig:move_map}. The accumulated resonance angle $\phi_{a}$ is calculated following the same procedure as before. Using the estimate of the deviation from 9:7 resonance at the point of entry given by Equation (\ref{X22}) with $e_{1}=0.02$, the value of $\phi_{a}$ is measured when semi-major axis ratio reaches 1.1829. As the results shown in Figure~\ref{fig:move_map} indicate, $\phi_{a}$ in the ``capture'' cases is also distributed in particular regions, which are indicated by green stripes. 
One of them remains close to $\phi_{\rm entry} = \frac{3}{2}\pi$ which is represented by dashed lines in this figure. 
We note that at least two narrow regions for capture are found in the interval  $[2\pi, 4\pi]$, which is similar to the situation with the simulations with initial $e_{1}=0.015$ and $e_{2}=0$.

In order to reveal the influence of the choice of initial eccentricities on the regions of capture, we show the results from three groups of simulations with different initial eccentricities in the ($\dot{a}/a$, $\phi_{a}$) plane in Figure~\ref{fig:ecc10_move}. The filled circles represent ``capture'' cases, and the open circles represent the ``fail'' cases. For the simulations with initial $e_{1}=0.005$ and $e_{2}=0$, the circles and capture regions are indicated by the violet color. Another two groups of simulations with higher initial eccentricities are illustrated  with  red and green colors as before. The positions of $\phi_{a}$ with $\phi_{\rm entry}=\frac{3}{2}\pi$ are also shown in the figure. In general, $\phi_{a}$ increases with smaller relative migration rate. For the results illustrated with the violet color, a scatter in $\phi_{a}$ appears for a fixed relative migration rate once this is} less than 2 ${\rm Myr}^{-1}$, which indicates that the influence of the circularization rate on the evolution of $\phi_{1}$ is playing a role in this region. On the other hand, the regions of capture for simulations with different initial eccentricities are located in different regions in the figure. In the region of small  $\phi_{a},$ ``capture'' is seen to occur at high relative migration rates; no violet stripes but several red and green stripes are found in that region. It is inferred that for simulations with high relative migration rate, larger initial eccentricities may lead to a higher probability of the system becoming locked in the 9:7 MMR. We also note that the distance between neighboring violet regions for capture is larger than for the corresponding green and red ones.

As mentioned above, planets with low initial eccentricities can be captured in 9:7 resonance with $\phi_{\rm entry}$ in only one window, while for planets with high initial eccentricities there are several discrete capture windows in the full $2\pi$ range of $\phi_{1}$.

\begin{figure}[htb!]
\includegraphics[width=\columnwidth]{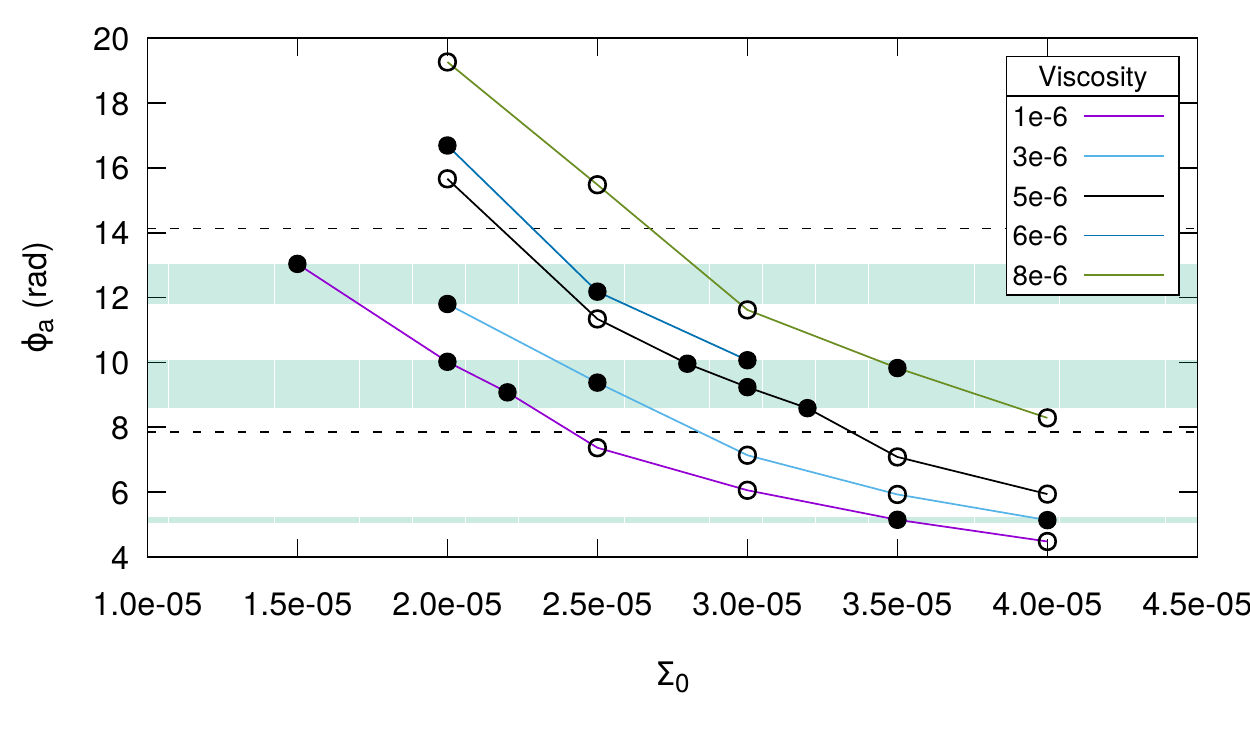}
\caption{Accumulated resonance angle $\phi_{a}$ as a function of surface density scaling parameter $\Sigma_0$ and viscosity $\nu$ in the simulations with initial $e_{1}=0.025$ and $e_{2}=0.015$. The filled and open circles show the ``capture'' cases and ``fail'' cases, respectively. 
Dashed lines indicate the position of $\phi_{a}$ when $\phi_{
\rm entry}=\frac{3}{2}\pi$. Green regions represent domains of $\phi_{a}$ for capture.
}
\label{fig:move_map}
\end{figure}

\begin{figure}[htb!]
\includegraphics[width=\columnwidth]{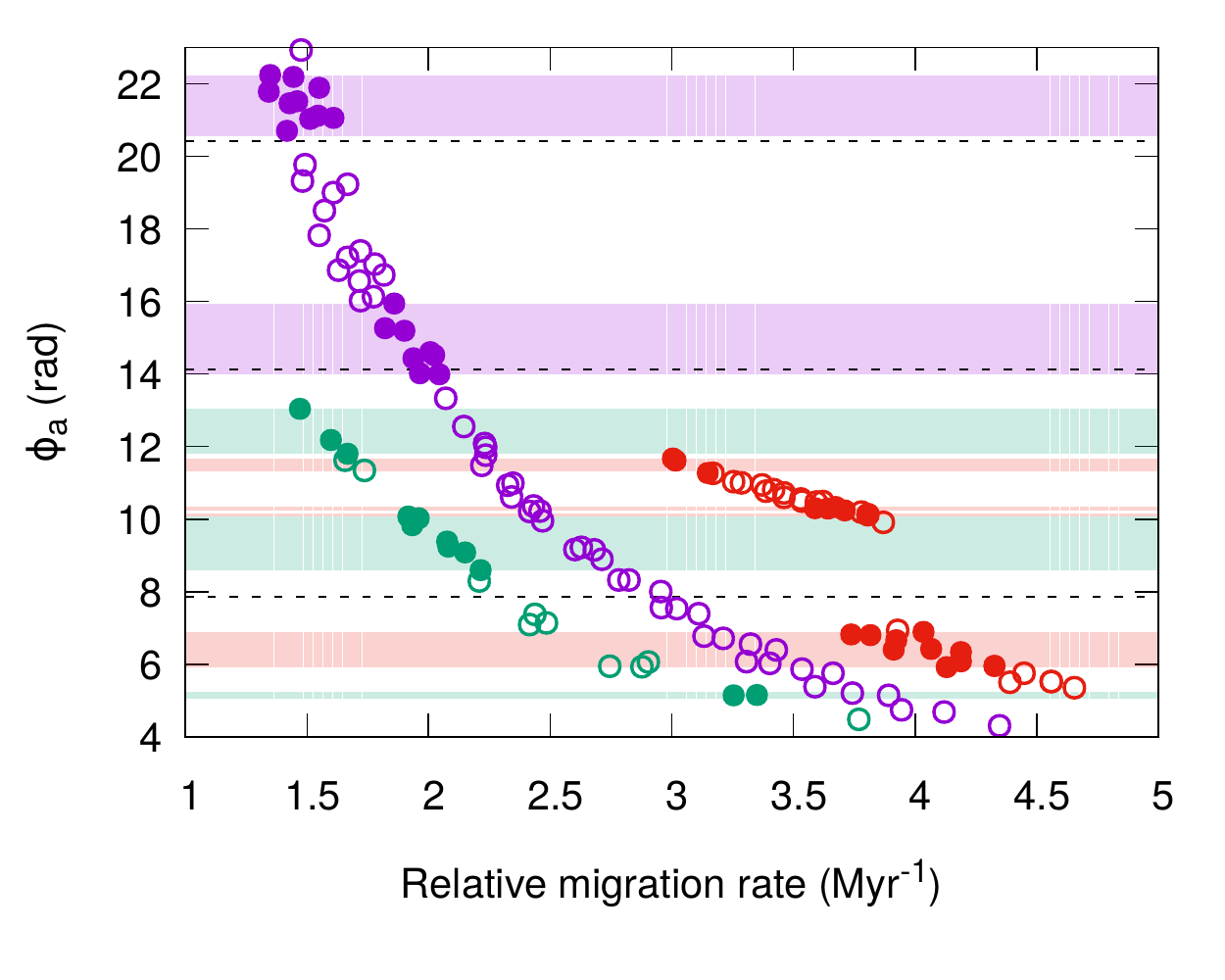}
\caption{Relation between $\phi_{a}$ and the initial relative migration rate in the simulations. The filled circles indicate the ``capture'' cases, and the open circles indicate the ``fail'' cases. The results of simulations with initial $e_{1}=0.005$ and $e_{2}=0$ are indicated by violet color. Red circles present the cases in which initial $e_{1}=0.015$ and $e_{2}=0$. Green circles represent the simulations that start with $e_{1}=0.025$ and $e_{2}=0.015$. The capture regions for each group of simulations are indicated with the same color-coding. The values of $\phi_{a}$ for which $\phi_{
\rm entry}=\frac{3}{2}\pi$ are indicated by dashed lines.}
\label{fig:ecc10_move}
\end{figure}


\section{Conclusions} \label{sec:conclusion}

We have performed hydrodynamic simulations of two low-mass planets ($m_{1}=m_{2}=3M_{\oplus}$) undergoing migration in a gaseous protoplanetary disk in order to study and verify the possibility of the capture of the system into 9:7 resonance. The disk models we considered were parameterized by a parameter, $\Sigma_0$ that scaled the surface density profile and the kinematic viscosity $\nu$. Practical considerations concerning the physical circumstances of the simulations, such as the migration and circularization rates of interest, meant that the systems were only followed
while they moved through a relatively small region of the protoplanetary disk.
 
For prescribed sets of initial conditions for the planetary orbits, we identified values of $\Sigma_0$ and $\nu$ that favored the formation of  a 9:7 mean motion commensurability through mapping simulations (see Section \ref{sec:survey}). From our survey, we confirm the general picture of the 9:7 resonant capture presented in previous works \citep[e.g.][]{Quillen2006, Mustill2011, Miga2017, Xu2017} and gave a description of this phenomenon as being related to one of the resonance angles having to be located in a particular range as the resonance is entered.

Using the mapping simulations with initial $e_{1}=0.005$ and $e_{2}=0$, we have demonstrated the existence of capture regions in $(\Sigma_{0}, \nu)$ parameter space in which 9:7 resonant capture is able to occur (see Figure~\ref{fig:ecc10_map}). We found that the system can become locked into the resonance when $\phi_{a}$ lies in a particular range (the capture window) and the width of this range depends on dynamical parameters (such as the relative migration rate and the circularization rate) that can be viewed as being determined by $\Sigma_{0}$ and $\nu$ (see Figure~\ref{fig:dot_angle}). A similar result has also been obtained in the work of \citet {Folonier2014}, which shows that the occurrence of capture into the 3:1 resonance depends on the initial values of the resonant angle and $\Delta \varpi$, in a system consisting of Jupiter and a small asteroid.

On the issue of the criterion for capture to take place in our survey, we find consistency with the result shown in Figure 2 of \citet {Mustill2011}, which illustrates the capture probability for a restricted three-body problem depending on the migration rate and the initial eccentricity. In our runs, the distribution of ``capture'' and ``fail'' cases corresponds to their result in the rescaled $(e_{1}$, $\dot{a}/a)$ plane (see Figure~\ref{fig:ecc10_comparison} in Section~\ref{sec:formation}).

We also have shown by rerunning simulations after slightly changing the initial semi-major axis of the outer planet how the 9:7 resonant capture depends on the initial orbits of the system (see Section \ref{sec:initial-orbit}).
This type of change does not significantly affect the dynamical parameters of the calculation, but, on account of the change in the time required to reach the resonance, there is expected to be a change in the value of the resonant angle $\phi_{1}$ when entering into the resonance, which may lead the system to behave in a different manner with regard to the window for resonant capture (see Figure~\ref{fig:ecc10_ini}). A similar conclusion was obtained by \citet {Miga2017} through {\it N}-body simulations.

In addition, in Section \ref{sec:ecc}, we studied the effect of the choice of the initial eccentricities on the window for $\phi_{1}$ that enable resonant capture by considering two groups of simulations. The first had $e_{1}=0.015$, $e_{2}=0$, and the second had $e_{1}=0.025$, $e_{2}=0.015$. 
The windows in these two cases were found to be  distributed over the full $2\pi$ region (see Figs.~\ref{fig:ecc9_region} and \ref{fig:move_map}). 
 
Compared with results from the mapping simulations for which $e_{1}=0.005$ and $e_{2}=0$, we found that for high relative migration rates in the range of ($3.0  -  4.5 ~ {\rm Myr}^{-1}$), the system can be locked in the 9:7 resonance when the initial eccentricities are higher (see Figure~\ref{fig:ecc10_move}). A similar result can be found in Figure 2 of \citet {Mustill2011} in which the probability of second-order resonance capture in the restricted three-body problem has a peak in the region of high eccentricities when the migration rate is very high.

In this paper we  have studied the formation of the 9:7 MMR in a system with two equal-mass super-Earths. The migration of the planets in the protoplanetary disk in our simulations is in the regime of type I migration. For our choice of the disk parameters, the planets are not able to open partial gaps. Other effects such as density wave propagation or wake-planet interactions that may prevent the resonant capture do not arise.
The formation of second-order resonances in a system of super-Earths moving over a larger radial extent in disks with a wider range of physical parameters will be the subject of future studies.

\acknowledgments

We thank the referee for a careful reading of the manuscript and a helpful report. We also would like to thank Cezary Migaszewski and Krzysztof Go{\'z}dziewski for the stimulating discussions about the second-order MMRS. We are indebted to Franco Ferrari for his continuous support in the development of our computational techniques and computer facilities.
J.C.B.P. thanks the Faculty of Mathematics and Physics, University of Szczecin for hospitality. 
We would like to acknowledge the support by Polish National Science Center MAESTRO grant DEC-2012/06/A/ST9/00276. The simulations were performed on HPC cluster HAL9000 of the Computing Center of the Faculty of Mathematics and Physics at the University of Szczecin.

\end{document}